\documentclass[10pt]{article}

\usepackage{amssymb}
\usepackage{amsmath}
\usepackage{multirow}
\usepackage{subfig}
\usepackage{color}
\usepackage[usenames,dvipsnames,svgnames,table]{xcolor}
\usepackage{soul}
\usepackage{graphicx}
\usepackage{authblk}

\newcommand{\be}{\begin{equation}} 
\newcommand{\ee}{\end{equation}}
\newcommand{\p}{\partial}
\newcommand{\grad}{\nabla}
\renewcommand{\v}{\mathbf{ v}}
\newcommand{\uvec}{\mathbf{ u}}

\newcommand{\fatf}{\mathbf{f}}
\renewcommand{\b}{\mathbf{ b}}
\newcommand{\A}{\mathbf{ A}}

\newcommand{\g}{\mathbf{ g}}
\newcommand{\res}{\mathbf{ r}}
\newcommand{\w}{\mathbf{ w}}
\newcommand{\n}{\mathbf{ n}}
\newcommand{\fgref}[1]{Fig. \ref{#1}}
\newcommand{\tbref}[1]{Table \ref{#1}}

\def\veps{\varepsilon}
\def\fatI{\mathbf{I}}
\def\fatsigma{\boldsymbol{\sigma}}
\def\fattau{\boldsymbol{\tau}}

\title{Dynamically coupling the non-linear Stokes equations with the Shallow Ice Approximation in glaciology: Description and first applications of the ISCAL method}

\author[1]{Josefin Ahlkrona}
\author[1]{Per L\"{o}tstedt}
\author[2,3]{Nina Kirchner}
 \author[4]{Thomas Zwinger}

\affil{Division of Scientific Computing, Department of Information Technology, Uppsala University, Uppsala, Sweden}
\affil{Department of Physical Geography, Stockholm University, Stockholm, Sweden}
\affil{Bolin Centre for Climate Research, Stockholm University, Stockholm, Sweden}
\affil{CSC - IT Center for Science Ltd., Espoo, Finland}

\begin{document}

\maketitle

\begin{abstract}
We propose and implement a new method, called the Ice Sheet Coupled Approximation Levels (ISCAL) method, for simulation of ice sheet flow in large domains under long time-intervals. The method couples the exact, full Stokes (FS) equations with the Shallow Ice Approximation (SIA). The part of the domain where SIA is applied is determined automatically and dynamically based on estimates of the modeling error. For a three dimensional model problem where the number of degrees of freedom is comparable to a real world application, ISCAL performs almost an order of magnitude faster with a low reduction in accuracy compared to a monolithic FS. Furthermore, ISCAL is shown to be able to detect rapid dynamic changes in the flow. Three different error estimations are applied and compared. Finally, ISCAL is applied to the Greenland Ice Sheet, proving ISCAL to be a potential valuable tool for the ice sheet modeling community.
\end{abstract}


\section{Introduction}  
With the increasing awareness of climate change and the associated sea level rise, research on ice sheets has intensified \cite{Alley, Hanna13, PollardDeConto, Shepherd, Vaughan}. Accurate predictions of the response of the Greenland Ice Sheet and the Antarctic Ice Sheet to global warming call for accurate numerical models, with an adequate representation of ice dynamics. The most accurate models of ice dynamics available to date, are so-called Full Stokes models (hereafter referred to as FS). In these models the ice dynamics are represented by the Stokes equations, treating ice as a slowly creeping, non-Newtonian, viscous fluid \cite{GreveBlatterBok}. These equations are computationally demanding to solve since they are highly non-linear, with a viscosity which depends on both strain-rate and temperature.

Computer performance restricts ice sheet simulations with FS to rather short timescales on the order of a few hundred years even on the largest parallel computers, see \cite{Hakime2012, Fabien2012, ISSM, Petra12, Petra14, ZJGRP}. Yet, the overall very slow response of ice sheets to external forcings (typical response times are thousands of years) calls for investigations of ice dynamics on much longer timescales. Longer simulations allows us to determine whether observed short-term fluctuations represent natural variability or are part of a long-term trend. Furthermore, we can validate ice sheet models by comparison to geological data, which provide a record of the long-term evolution of ice sheets that took place before the beginning of the modern, mostly satellite-based, observational period. Such simulations of long-term ice sheet behavior are known as paleo-simulations, which typically also involve considerably larger spatial domains than those currently covered. These simulations are not yet feasible with FS.

Numerical simulations of ice sheet behavior during glacial cycles at hemispheric scales are possible \cite{Siegert,Colleoni,Tarasov,Forsstrom} if an approximation to the Stokes equations is used, known as the Shallow Ice Approximation (hereafter referred to as SIA) \cite{BlaHaftet, FowLar, Hindmarsh}. The SIA is based on the assumption that the ice body is shallow, and that vertical shearing motion dominates over horizontal stretching and shearing. This implies substantial simplifications to the FS equations, thereby rendering SIA models computationally very cheap.

The overall, inland, ice sheet behavior is often modeled sufficiently well with SIA, but is poorly represented in regions which are now regarded as critical for the mass budget of ice sheets and for future sea level rise. Such regions are the highly dynamic ice sheet margin, where ice streams (faster flowing regions within the ice) terminate at the ocean. In the ocean, the ice either breaks into icebergs, or forms large floating ice platforms which remain connected to the ice sheet, called ice shelves. At ice shelves, and at domes or over rapidly undulating basal terrain, an accurate representation of ice dynamics is also not possible in SIA \cite{BlaHaftet,SchoofHindmarsh,Hutter83,Ahlkrona13,Ahlkrona13a}. As a consequence of approximations being applicable in some regions of an ice sheet, and in some other not, the idea of coupling the FS with approximations emerged in the ice sheet modeling community, presented first by Seroussi et al \cite{Seroussi12} in 2012. In \cite{Seroussi12}, a coupling between the so-called \textit{Blatter-Pattyn approximation} \cite{Blatter95,Pattyn2003}, the \textit{Shelfy Stream Approximation} \cite{SSA}, and FS was demonstrated for an idealized floating ice shelf.

Theoretical analyses and model case studies have been devoted to understand precisely under what conditions and in which parts of an ice sheet SIA can be used and is expected to give accurate results \cite{Hindmarsh, SchoofHindmarsh,Ahlkrona13, ElmerVSsia, GagliardiniZwinger}. Yet, it is not easy to transfer such insights to simulations of real ice sheets, featuring complicated geometry over complex topography, spatio-temporally variable basal conditions and climate forcings. It is practically impossible to determine {\it a priori} when, and in which parts of a sheet, the SIA is applicable.

Here, we propose a new method, called the Ice Sheet Coupled Approximation Level (hereafter referred to as ISCAL). This method automatically and dynamically decides from within the simulation where in an ice sheet the SIA is a valid approximation, and applies the FS elsewhere. The ISCAL method is computationally less expensive than FS, and more accurate than the SIA. The ISCAL method goes beyond the method in \cite{Seroussi12}, providing an automatic switch between FS and SIA based on estimated modeling errors, and coupling the FS and the SIA in a dynamic manner rather than statically. ISCAL has the potential to simulate ice sheet dynamics on larger spatial and temporal domains that were previously considered inaccessible with FS alone. This opens the way to more accurate paleo-simulations, model validation by comparison to geological archives, and improved assessment of observed ice dynamics. ISCAL can also speed up simulations on shorter time intervals for parameter sensitivity studies or spin-up simulations to determine consistent initial conditions for FS where the accuracy requirements are lower. 

The aim of this paper is to present the ISCAL algorithm, which is implemented in the finite element software Elmer/Ice \cite{ElmerDescrip}, and to evaluate its properties. In Section~\ref{sec:FSSIA}, the theoretical framework of FS and SIA is briefly reviewed. In Section~\ref{sec:nummeth}, the dynamic coupling of FS and SIA, and the procedureof how to estimate where the SIA is valid to a given accuracy, is described. In Section~\ref{sec:numexp}, we perform numerical experiments to demonstrate the accuracy, efficiency and flexibility of ISCAL using a transient conceptual model problem in three dimensions (3D). We also apply ISCAL to real data from the Greenland Ice Sheet to prove the maturity of the method and its potential value to the glaciological community.

\section{Theory of FS and SIA}\label{sec:FSSIA}

On time scales relevant for ice sheet dynamics, ice can be described as an isotropic, viscous, incompressible, non-Newtonian fluid, with a very low Reynolds number. The dynamics is governed by the non-linear Stokes equations, \textit{i.e.} the FS equations. The SIA is derived from the FS equations by assuming that the aspect ratio, \textit{i.e.} the ratio $\veps= [H]/[L]$ of a typical thickness of the ice $[H]$ to a typical lateral length scale $[L]$ is small \cite{BlaHaftet,Ahlkrona13,Ahlkrona13a}. By expanding the solution in the small parameter and solving for the lowest order terms, the SIA is obtained (see \cite{BlaHaftet,Hutter83}). In this section, the FS and SIA equations are described. We consider isothermal conditions for simplicity, although ISCAL is implemented for a fully thermo-dynamically coupled setting. A partial derivative of a variable $u$ with respect to an independent variable $x$ is written $\partial_{x}u$.
\begin{figure}[h]
    \centering
   \includegraphics[width=0.8\textwidth]{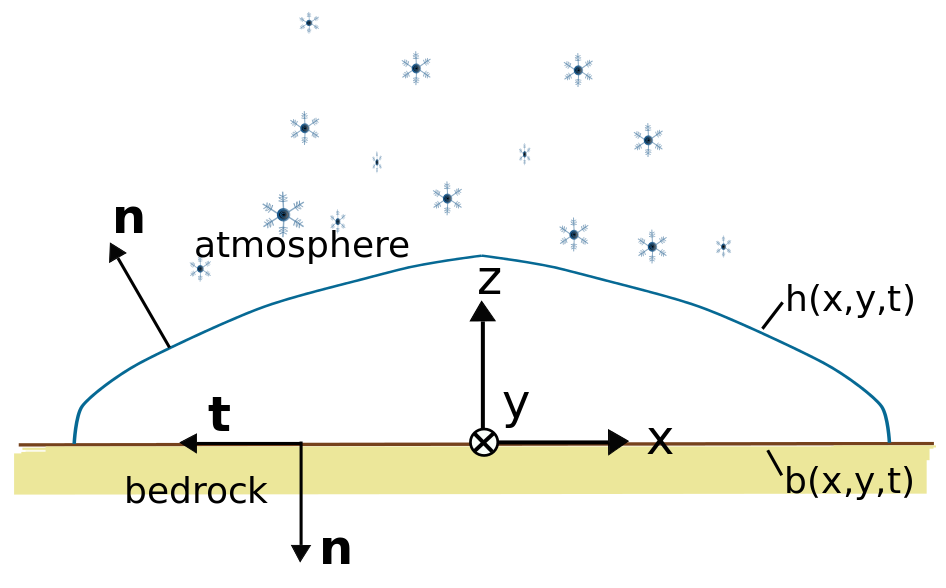}
   \caption{An ice-sheet in a Cartesian coordinate system, exaggerated in the $z$-direction. The ice surface position is described by $h(x,y,t)$ and the bedrock underneath is given by $b(x,y,t)$. The normal vector, $\mathbf{n}$ is pointing outwards from the ice. In two dimensions there is one tangential vector $\mathbf{t}$, and in 3D we choose two that span a tangential plane.} 
   \label{fig:coords}
\end{figure}

The general setting is as follows: We consider a grounded ice sheet in a Cartesian coordinate system, see \fgref{fig:coords}, where the ice sheet surface $z=h(x,y,t)$ is a function of position $(x,y)$ and time $t$, and the ice sheet base is $z=b(x,y,t)$. The ice sheet surface evolves as
\be
\label{eq:freesurface}
{\p_t h}+v_x|_{z=h} {\p_x h}+v_y|_{z=h}{\p_y h}-v_z=a_s,
\ee
where $a_s=a_s(x,y,t)$ is a prescribed source term denoting the net accumulation normal to the surface of the ice sheet, and depends on the climate conditions. The velocity field $\v=(v_x, v_y, v_z)^T$ is obtained as the solution to the FS or SIA equations.

The bedrock underneath the ice is here assumed to be rigid, so that $b(x,y,t)=b(x,y)$. For simulations covering long time spans, an evolution equation for $b$ similar to \eqref{eq:freesurface} is solved to account for isostatic adjustment and bottom melting/refreezing. At the boundaries (at \textit{i.e.} the ice surface and at the ice sheet base), a normal unit vector ${\bf n}$ points away from the ice body, and two vectors ${\bf t}_1 $ and ${\bf t}_2 $ span a tangential plane.

\subsection{The full Stokes formulation}
The FS equations that are solved for the velocity $\v=(v_x,v_y,v_z)^T$ and the pressure $p$ are
\begin{subequations}
\label{eq:vFS} 
\begin{alignat}{2}
\label{eq:momentum}
-\grad p +\nabla\cdot\left( \eta (\grad \v + (\grad \v)^T) \right)  + \rho \g &= \textbf{0}, \\
\label{eq:mass} \nabla\cdot \v &= 0,
\end{alignat}
\end{subequations}
where $\rho$ is the density, and $\rho\g$ is the force of gravity. The viscosity, $\eta$, depends on the velocity as 
\be\label{eq:constitutivelaw}
\eta  = \frac{1}{2}\mathcal{A}^{-1/n}d^{-(1-1/n)}, 
\ee
The Glen exponent $n$ is usually chosen as $n=3$. The rate factor $\mathcal{A}=\mathcal{A}(T')$ is coupling the pressure melting point corrected temperature field, $T'$, to the velocity field through an exponential function \cite{GreveBlatterBok}. Due to convection in the temperature equation, there is thus a two-way coupling between temperature and velocity. As we only consider isothermal conditions, we will not describe this here, and the value of $\mathcal{A}$ is simply a constant given in Section \ref{sec:modeldescrip}. The effective strain rate, $d$, is given by
\be \label{eq:eISCALtrain}
\begin{array}{rl}
d=& \sqrt{\left(\frac{1}{2}{\rm tr}\left(\frac{1}{2}\left(\grad \v+ (\grad \v)^T \right)^2\right)\right)}
= \frac{1}{\sqrt{2}}\left(\left({\partial_x v_x}\right)^2+\left({\partial_y v_y}\right)^2 +\left({\partial_z v_z}\right)^2\right. \\
&+ \left.\frac{1}{2} \left({\partial_y v_x}+{\partial_x v_y}\right)^2 +\frac{1}{2} \left({\partial_z v_x}+{\partial_x v_z}\right)^2 +\frac{1}{2} \left({\partial v_y}+{\partial_y v_z}\right)^2  \right)^{1/2} \, .
\end{array} 
\ee
Since typically $n>1$, there is a singularity obtained in \eqref{eq:constitutivelaw} at $d=0$, which can cause numerical difficulties. To regularize the problem, the shear rate is bounded below by a small number, $d_0$ \cite{ElmerManual}. 

Due to the velocity dependent viscosity in \eqref{eq:constitutivelaw}, the problem \eqref{eq:vFS} is highly non-linear. Becasue of this non-linearity, and due to the fact that the equation system in \eqref{eq:vFS} is a saddle-point problem, the FS equations are computationally expensive to solve.

\subsection{Boundary conditions}

At the ice surface, the atmospheric stresses are generally considered to be negligible implying
\be\label{eq:bndcondtop}
  \fatsigma \cdot \n=0
\ee
The Cauchy stress $\fatsigma$ is related to the deviatoric stress tensor $\fattau$, velocity and pressure by 
\be
\fatsigma=-p\fatI +\fattau=-p\fatI +\eta (\grad \v + (\grad \v)^T), 
\ee
\textit{i.e.}, at the surface we have a homogeneous Neumann condition for the velocity field and a homogeneous Dirichlet condition for the pressure. 
At the ice-bedrock interface, it is less obvious how to choose boundary conditions. The conditions there depend on geological, hydrological and thermal conditions which are often inaccessible to observations. We employ two different boundary conditions for the velocity $\v$ at the base (indicated by subscript $b$): a no slip condition 
\be\label{eq:bndslip}
  \v|_b=0, 
\ee
which is accurate if the ice base is frozen, or a linear sliding law, relating the basal velocities with basal shear stress as
\begin{subequations} \label{eq:slidinglaw}
\begin{alignat}{2}
 \label{eq:slidinglawt}(\v \cdot \mathbf{t_i}) |_b &= -\tau_b/\beta = -(\mathbf{t_i} \cdot {\fatsigma}\cdot \n) |_b /\beta, \quad i=1,2\\
 \label{eq:slidinglawn}(\v \cdot \mathbf{n})|_b &= 0.
\end{alignat}
\end{subequations}
The vectors $t_1$ and $t_2$ span a tangential plane, and the normal vector $\mathbf{n}$ is pointing downwards (\fgref{fig:coords}). We assume that no melting or refreezing occurs at the bed, hence \eqref{eq:slidinglawn} holds. The friction between the ice and the underlying bedrock is described by the sliding coefficient $\beta$, which will be small in high sliding areas such as under ice streams/outlet and glaciers.

\subsection{The Shallow Ice Approximation}
Due to the approximations made in deriving SIA, the SIA velocities and pressure are very simple to compute and are directly given by:
\begin{subequations}
\label{eq:vSIA} 
\begin{alignat}{4}
\label{eq:vxSIA}v_{x}&=v_{b,x}-2(\rho g)^n {\p_x h}||\grad h||^{n-1}\int_b^z\mathcal{A}(T')(h-z')^ndz',\\
\label{eq:vySIA}v_{y}&=v_{b,y}-2(\rho g)^n {\p_y h}||\grad h||^{n-1}\int_b^z\mathcal{A}(T')(h-z')^ndz', \\
\label{eq:vzSIA}v_{z}&=v_{b,z}-\int_b^z\left({\p_xv_x}+{\p_y v_y}\right)dz', \\
\label{eq:pSIA}p&=\rho g (h-z),
\end{alignat}
\end{subequations}
see \cite{GreveBlatterBok}. The norms in \eqref{eq:vSIA} and in all following equations is the Euclidean vector norm. To arrive at the integral forms of \eqref{eq:vSIA}, the boundary conditions were used, and except for a sliding law and a melting/refreezing model describing the basal velocity $(v_{b,x},v_{b,y},v_{b,z})$, no further equations are needed. The sliding law is the linear sliding law in \eqref{eq:slidinglaw}, but the Cauchy stress tensor at the base, $\fatsigma$, is approximated as
\begin{equation}\label{eq:siasigma}
\fatsigma=\left(\begin{smallmatrix}
-\rho g (h-b)  & 0  & -\rho g\partial_xh(h-b) \\ 
0 & -\rho g (h-b) &  -\rho g\partial_yh(h-b) \\ 
 -\rho g\partial_xh(h-b)& -\rho g\partial_yh(h-b) & -\rho g (h-b) 
\end{smallmatrix}\right).
\end{equation}
As in \eqref{eq:slidinglawn}, we assume there is no melting or refreezing at the ice base.

We previously examined the validity and accuracy of the SIA in \cite{Ahlkrona13} and \cite{Ahlkrona13a} by analysis and numerical simulations of 2D ice flow over a bumpy bed. As expected, the accuracy of the SIA decreases when the aspect ratio $\veps$ increases. For aspect ratios above $10^{-1}$ the relative error is of $\mathcal{O}(1)$. Furthermore, the SIA theory assumes that the surface slope is proportional to $\veps$, \cite{Ahlkrona13a}, that shearing dominates sliding, and that horizontal in relation to vertical velocities scale with the aspect ratio $\epsilon$ \cite{SchoofHindmarsh}. The SIA is thus not expected to perform well in regions with large spatial variations in data (\textit{e.g.} topography), at steep margins, in ice streams or ice shelves, and at domes.

\subsection{Time evolution with FS and SIA}

The FS and SIA equations \eqref{eq:vFS} and \eqref{eq:vSIA} are time-independent. Time-dependency enters in the  free surface problem in \eqref{eq:freesurface}, and in the temperature equation (not considered here), which in turn determines the evolution of the viscosity. The surface evolution equation \eqref{eq:freesurface} exists in both a non-approximated version \eqref{eq:freesurface} and an approximated SIA-version. The computational gain in approximating them with SIA is far less than for the solution of the Stokes equations. In our simulations, we stick to the solver already implemented in Elmer/Ice \cite{ElmerDescrip}. The following discussion concerning the ISCAL method will focus on the computation of the velocity and pressure fields.

\section{Numerical method}\label{sec:nummeth}
The numerical solution of the ice sheet evolution consists of two parts. First, the FS and SIA equations \eqref{eq:vFS} and \eqref{eq:vSIA} are solved for the velocity $\v$ with a fixed ice domain and a computational mesh covering the domain at time $t$. Then a new height at $t+\Delta t$ is computed by integrating \eqref{eq:freesurface} in time using the velocity computed at $t$. The mesh is then adjusted in the vertical direction to account for the elevation changes and \eqref{eq:vFS} and \eqref{eq:vSIA} are solved again at $t+\Delta t$. The backward Euler time integration is first order accurate with an error of $\mathcal{O}(\Delta t)$.

Since ISCAL is based on the FS and SIA equations, we first briefly describe how to solve the respective equations and then proceed to the combination in the ISCAL method in more detail. 

\subsection{Solving the full Stokes equations}
\begin{figure}
    \centering
    \includegraphics[trim={1cm 6cm 1.5cm 5cm},clip,width=0.6\textwidth]{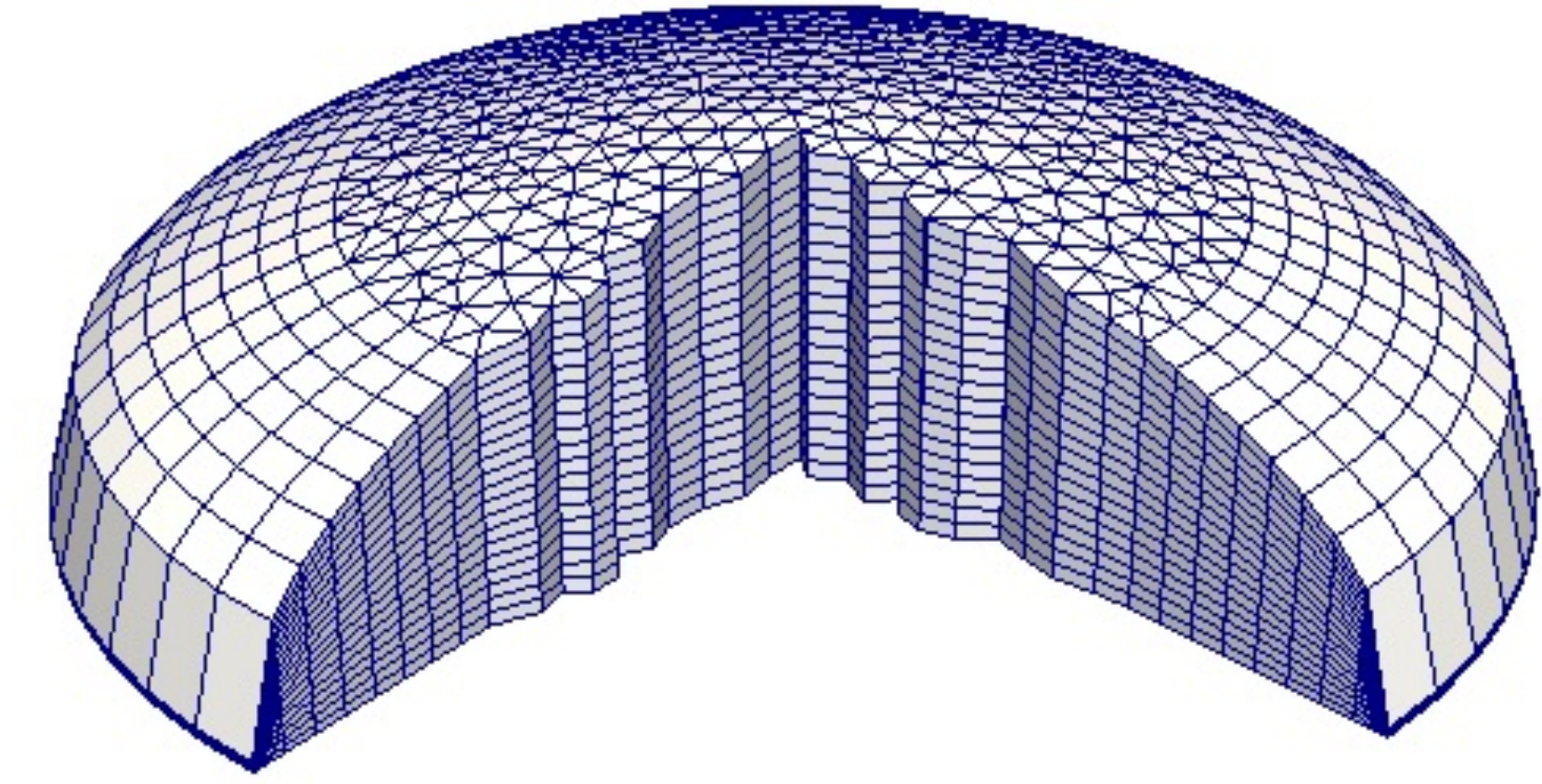}
    \caption{An extruded, 3D mesh on a circular ice sheet, with 17860 nodes, distributed in $20$ vertical layers. The vertical axis in the graph is scaled by a factor 100 in the figure. A slice is cut out to show how the nodes are aligned in the vertical direction. This 2D footprint mesh is unstructured in the center and structured at the margins.}%
    \label{fig:ExtrudedMesh}
\end{figure}
Finite element discretizations of the FS equations, the ice surface evolution, and the temperature evolution are already implemented in Elmer/Ice \cite{ElmerDescrip}. The mesh consists of triangles or quadrilaterals in the $x-y$ plane extruded in the $z$ direction to the ice surface to form prisms, see \fgref{fig:ExtrudedMesh}. This type of mesh is favorable for ice sheet modeling. Linear elements are used, and the discretization of \eqref{eq:vFS} with a given (\textit{e.g.} from a previous iteration) viscosity, $\eta$, leads to a system of linear equations
\be\label{eq:lineq}
   \A\uvec=\mathbf{b},
\ee 
where the components of the solution vector $\uvec$ are the nodal values of $\v$ and $p$. The right hand side, $\mathbf{b}$, is an expression of the gravitational force for the momentum equations. The solution of \eqref{eq:lineq} with a fixed $\A$ is computed by an iterative method. Because of the non-linear rheology of ice, the system matrix $\A$ depends on $\eta$ which in turn depends on the velocity, $\v$, through \eqref{eq:constitutivelaw}. Hence,  $\A=\A(\eta(\uvec))=\A(\uvec)$. Either a Picard iteration or a Newton iteration is applied to solve the non-linear equation \eqref{eq:lineq}, generating a sequence of solutions $\uvec_k,(\; k=1,2,\ldots)$. Both the linear system iterations and the non-linear iterations are considered as converged when the difference (measured in $||\cdot ||$) between two consecutive $\uvec_{k+1}$ and $\uvec_k$ is sufficiently small. The initial guess $\uvec_0$ in the non-linear iterations is the solution from the previous time step. This iteration in combination with large domains and long time intervals are the main reasons why ice sheet modeling is a computationally demanding problem. 

\subsection{Solving the SIA equations}

The SIA solution is determined by computing the right hand side expressions in \eqref{eq:vSIA}. As there are no partial differential equations to solve, the numerics is simple. The gradients in \eqref{eq:vSIA} are computed using the finite element framework of Elmer/Ice. As the meshes are constructed such that the nodes are aligned in the vertical direction, (\fgref{fig:ExtrudedMesh}), the integrals in \eqref{eq:vSIA} can  be computed using the trapezoidal rule. 

\subsection{Coupling FS and SIA}

\begin{figure}
    \centering
    \includegraphics[width=0.6\textwidth]{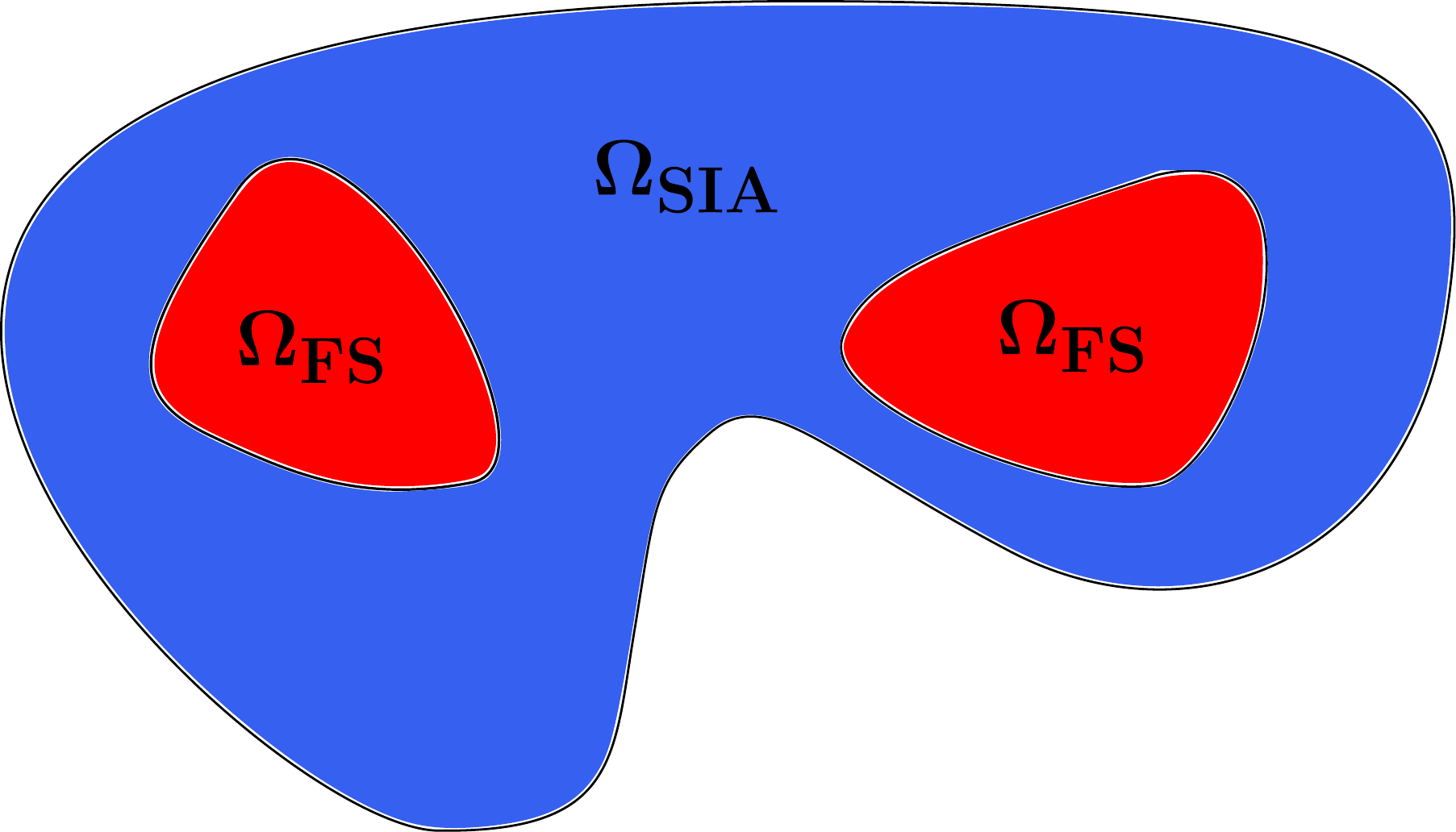}
    \caption{The domain $\Omega$ is automatically partitioned into subdomains $\Omega_{SIA}$ where the SIA is sufficiently accurate, and other subdomains $\Omega_{FS}$ where the FS equations need to be solved such that $\Omega=\Omega_{FS} \cup \Omega_{SIA}$.}
    \label{fig:Domains}
\end{figure}

Let us introduce some specific notation for describing the coupling. The SIA solution is denoted by $\uvec_{SIA}$, the FS solution by $\uvec_{FS}$, and we let the combined ISCAL solution, obtained by coupling FS and SIA, be written $\uvec_{ISCAL}$. The computational domain of the ice sheet, $\Omega$, is divided into a subdomain (or a collection of subdomains) $\Omega_{SIA}$, in which the SIA is sufficiently accurate, and in the complement $\Omega_{FS}=\Omega\backslash\Omega_{SIA}$ where the FS equations must be solved, see \fgref{fig:Domains}. The nodes in the solution vector are reordered according to which domain they belong. Then \eqref{eq:lineq} can be rewritten as a $2\times 2$-block system, one part representing the FS subdomains, and one part representing the SIA subdomains
\begin{equation}\label{eq:split}
\A\uvec_{FS}^{\Omega}=\left(\begin{matrix}
\A_{FS} &  \A_{CO}\\ 
\A_{OC} & \A_{SIA}
\end{matrix}\right)\left(\begin{matrix}
\uvec^{\Omega_{FS}}_{FS}\\ 
\uvec^{\Omega_{SIA}}_{FS}
\end{matrix}\right)=\left(\begin{matrix}
\b^{\Omega_{FS}}\\ 
\b^{\Omega_{SIA}}
\end{matrix}\right).
\end{equation}
In this equation, $\A_{FS}$ is the FS system matrix for the FS area, $\A_{SIA}$ corresponds to the SIA area, $\A_{CO}$ represents the dependency of the solution in the FS area on the values of $\uvec$ in the SIA area, and $\A_{OC}$ represents the dependency of the solution in the SIA area on the solution in the FS area. Both $\A_{FS}$ and $\A_{SIA}$ are square matrices. In the beginning of each time-step, $\uvec_{SIA}$ is computed for the whole domain $\Omega$ by evaluating the expressions in \eqref{eq:vSIA}. This computational work is negligible in comparison to solving the FS equations. The solution $\uvec_{FS}^{\Omega_{SIA}}$ can then be approximated by $\uvec_{SIA}^{\Omega_{SIA}}$ in $\Omega_{SIA}$, and we only have to solve for $\uvec^{\Omega_{FS}}_{FS}$, disregarding the entire second row in \eqref{eq:split}. The remaining equation in the top row of \eqref{eq:split} is then
\be\label{eq:AFSeq}
\A_{FS}\tilde{\uvec}^{\Omega_{FS}}_{FS}+\A_{CO}\uvec^{\Omega_{SIA}}_{SIA}=\b^{\Omega_{FS}}.
\ee
The solution $\tilde{\uvec}^{\Omega_{FS}}_{FS}$ is an approximation of ${\uvec}^{\Omega_{FS}}_{FS}$ since $\uvec^{\Omega_{SIA}}_{FS}$ is replaced  by $\uvec^{\Omega_{SIA}}_{SIA}$. The SIA solution $\uvec^{\Omega_{SIA}}_{SIA}$ is known in $\Omega$ and we can move the second term  to the right hand side in \eqref{eq:AFSeq} to arrive at an equation for $\tilde{\uvec}^{\Omega_{FS}}_{FS}$
\be\label{eq:ISCALsystem}
\A_{FS}\tilde{\uvec}^{\Omega_{FS}}_{FS}=\b^{\Omega_{FS}}-\A_{CO}\uvec^{\Omega_{SIA}}_{SIA} .
\ee
In this way, the SIA solution $\uvec_{SIA}^{\Omega_{SIA}}$ will act as Dirichlet boundary conditions for the FS solution at the interfaces between $\Omega_{FS}$ and $\Omega_{SIA}$. Remember that ice is a non-Newtonian fluid with a viscosity dependent on $\v$ such that $\A_{FS}=\A_{FS}(\tilde{\uvec}^{\Omega_{FS}}_{FS})$ and $\A_{CO}=\A_{CO}(\tilde{\uvec}^{\Omega_{FS}}_{FS})$. The system \eqref{eq:ISCALsystem} has to be solved iteratively by a non-linear Picard or Newton iteration, but $\tilde{\uvec}^{\Omega_{FS}}_{FS}$ has much fewer unknowns than $\uvec_{FS}^{\Omega}$ in \eqref{eq:split}. At convergence, the complete ISCAL solution is merged together as $\uvec_{ISCAL}^T=(\uvec^{\Omega}_{ISCAL})^T=((\tilde{\uvec}^{\Omega_{FS}}_{FS})^T,(\uvec^{\Omega_{SIA}}_{SIA})^T)$. 

\begin{figure}
    \centering
    \subfloat[FS]{{\includegraphics[width=0.35\textwidth]{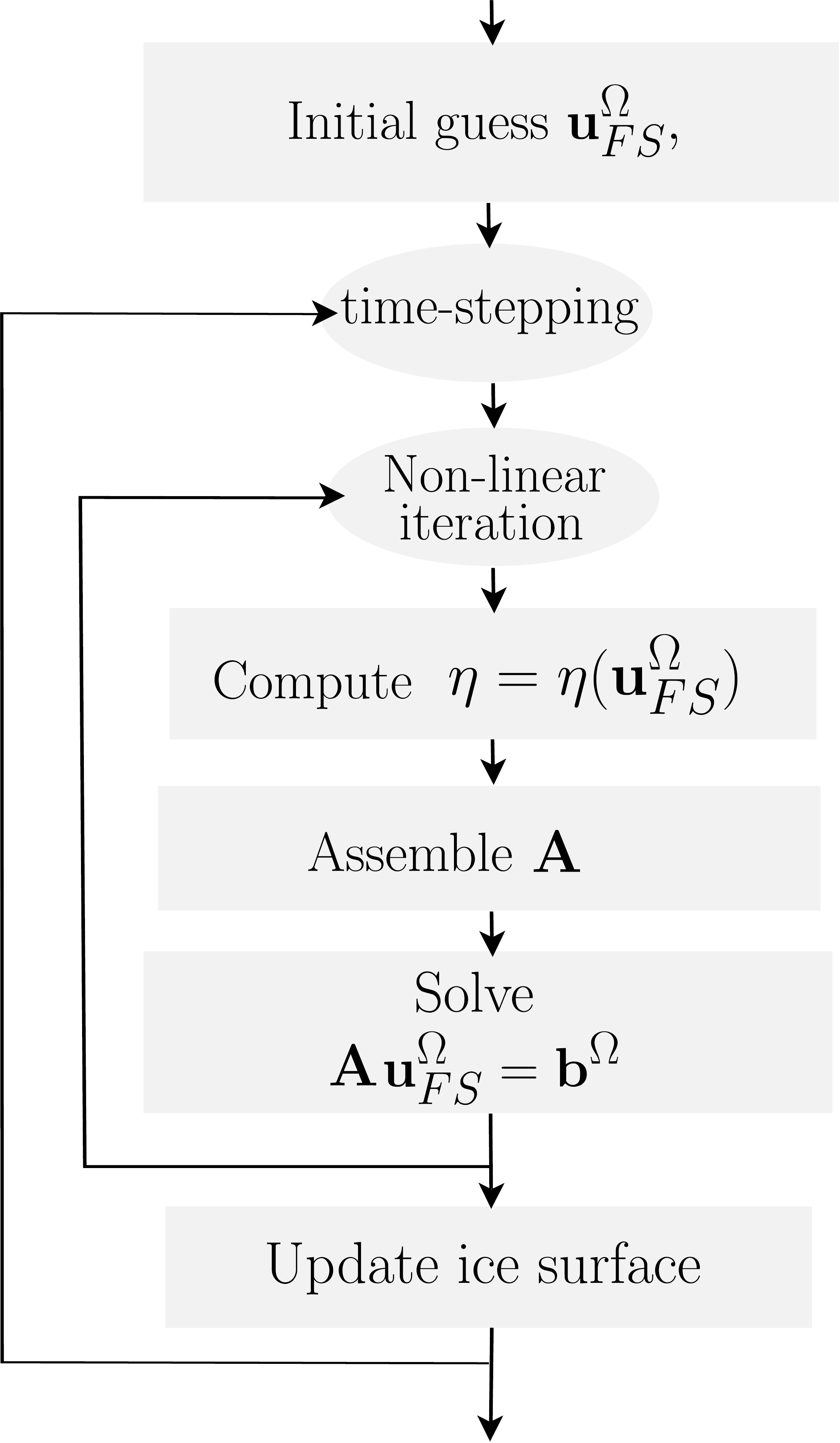} }}%
    \,
    \subfloat[ISCAL]{{\includegraphics[width=0.55\textwidth]{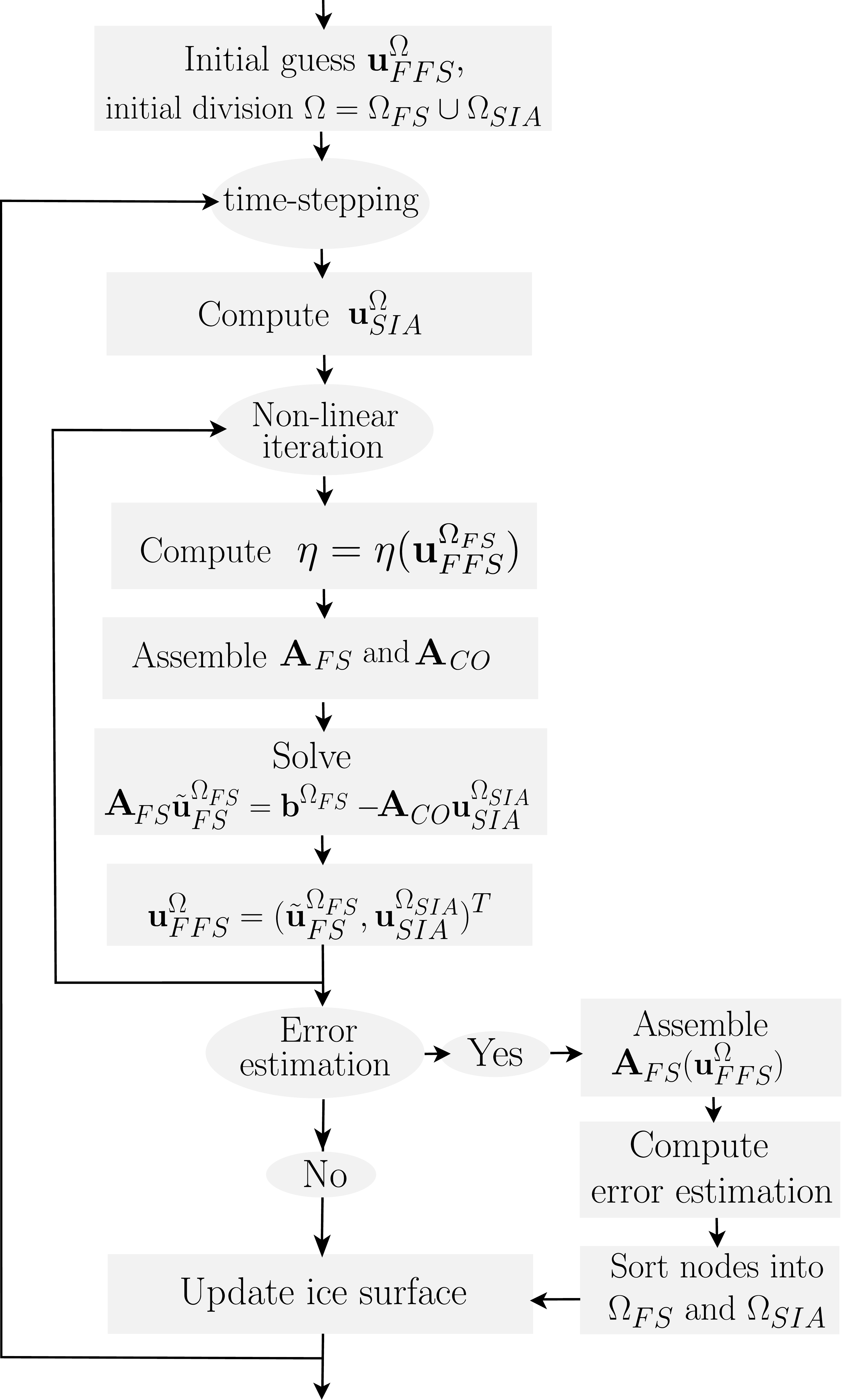} }}%
    \caption{Algorithms for computing ice sheet evolution by solving the FS equations and the ISCAL equations.}%
    \label{fig:Algorithm}%
\end{figure}

An outline of the algorithm is found in \fgref{fig:Algorithm}b. In comparison to the FS solver in \fgref{fig:Algorithm}a, error estimation, a SIA solver, and administration of the partitioning of $\Omega$ are added in the ISCAL solver.

\subsection{Model adaptivity and error estimation}\label{sec:ErrorEstimation}
It is difficult to know {\it a priori} exactly where in an ice sheet the SIA is a valid approximation, and the answer will change when the ice sheet evolves in time. Therefore, we use an automatic error estimation of a linearized problem, which determines in which parts of the ice sheet the SIA can be applied, and split the domain $\Omega$ into $\Omega_{SIA}$ and $\Omega_{FS}$ accordingly. The highest update frequency for the splitting of the domain is at most at the end of every time step, but for most problems this update can be much less frequent. Since there are several possible measures of the accuracy of the SIA, we construct three different error estimations; one based on the velocity, one on the residual of \eqref{eq:vFS}, and one on a functional of the velocity. The parameters for the error estimations, such as the error tolerance, $\epsilon$, and the number of timesteps there are between each error estimation, $m$, are user defined.

\subsubsection{Estimating the error in the velocity field}\label{sec:Error1}

Determining accurate ice velocities, \textit{e.g.} for estimating mass loss or for comparison with satellite data, is an important glaciological problem. From this viewpoint, it is of interest to control the error in the SIA velocity, \textit{i.e.} in the velocity components of 
\be\label{eq:trueerror}
\Delta \uvec^\Omega_{SIA} =\uvec^\Omega_{FS}-\uvec^\Omega_{SIA}.
\ee
Since we do not have access to $\uvec_{FS}$,  the error is approximated by comparing to a reference solution, $\uvec_{REF}$, instead, \textit{i.e.} 
\be\label{eq:estimatederror}
\tilde{\Delta \uvec}^\Omega_{SIA}=\uvec^\Omega_{REF}-\uvec^\Omega_{SIA}. 
\ee
The reference solution, $\uvec_{REF}$, is obtained once a solution $\uvec_{ISCAL}^{\Omega}$ has been found, by solving the linear system
\be\label{eq:Reference}
\tilde{\A}\uvec^{\Omega}_{REF}=\b,
\ee
where $\tilde{\A}=\A(\uvec^{\Omega}_{ISCAL})$. For comparison remember that the FS solution satisfies
\be\label{eq:FSeq}
{\A}(\uvec_{FS}^{\Omega})\uvec_{FS}^{\Omega}=\b.
\ee
In order for the error estimation to be accurate, \textit{i.e.} for $\tilde{\Delta \uvec}^\Omega_{SIA} \approx \Delta \uvec^\Omega_{SIA}$, the difference between $\uvec_{FS}^{\Omega}$ and $\uvec^{\Omega}_{REF}$ should be small. By denoting the error in $\uvec_{ISCAL}^{\Omega}$ by $\Delta \uvec_{ISCAL}$ such that
\be\label{eq:FSSerr}
\Delta \uvec_{ISCAL}=\uvec_{FS}^{\Omega}-\uvec_{ISCAL}^{\Omega},
\ee
and using \eqref{eq:FSeq} we get
\be\label{eq:FSeqexpand}
\begin{array}{ll}
{\A}(\uvec_{ISCAL}^{\Omega}+\Delta \uvec_{ISCAL})(\uvec_{ISCAL}^{\Omega}+\Delta \uvec_{ISCAL})\\
=(\tilde{\A}+ {\A}(\uvec_{ISCAL}^{\Omega}+\Delta \uvec_{ISCAL})-{\A}(\uvec_{ISCAL}^{\Omega}))(\uvec_{ISCAL}^{\Omega}+\Delta \uvec_{ISCAL})\\
=\tilde{\A}\uvec_{FS}^{\Omega}+\p\A\Delta\uvec_{ISCAL}\uvec_{FS}^{\Omega}=\b,
\end{array}
\ee
where an element $\p A_{ijk}$ of $\p\A$ is the first derivative of $A_{ij}$ with respect to $u_k$. From \eqref{eq:FSeqexpand} we conclude that if $\|\p\A\Delta\uvec_{ISCAL}\,\uvec_{FS}^{\Omega}\|$ is small then $\uvec^{\Omega}_{REF}$ in \eqref{eq:Reference} is a good approximation of $\uvec_{FS}^{\Omega}$ in \eqref{eq:FSSerr} and $\tilde{\Delta \uvec}^\Omega_{SIA} \approx \Delta \uvec^\Omega_{SIA}$.

The computation of $\uvec_{REF}^{\Omega}$ only requires the assembly and solution of a system of linear equations.  The extra computational work is acceptable, in particular since the error estimation is generally not necessary in every time step. The additional computational time attributed to the error estimation is measured in Section \ref{Efficiency}.

In an ice sheet, the major, most easily observed flow of ice is in the horizontal plane, and we therefore control the error only in the horizontal velocities $(v_x,v_y)=(u_{v_x},u_{v_y})$. It is, however, easy to adapt the code into estimating the error in all components of the solution. To determine whether SIA is sufficiently accurate in a node $i$ in the domain, both the relative and absolute differences between the nodal values $(v_{xi},v_{yi})_{SIA}$ and $(v_{xi},v_{yi})_{REF}$ are considered at each node ($i=1,\ldots, N$), such that if
\begin{equation}\label{eq:errorest1}
\begin{array}{ll}
\|(\tilde{\Delta u}_{v_{x}i} ,\tilde{\Delta u}_{v_{y}i} )_{SIA} \|= \|(v_{xi},v_{yi})_{SIA} -(v_{xi},v_{yi})_{REF}\| \\\leq \max\left(\epsilon_{rel}\|(v_{xi},v_{yi})_{REF}\|,\epsilon_{abs}\right)
\end{array}
\end{equation}
then the SIA is sufficiently accurate, and the node is included in $\Omega_{SIA}$. The absolute and relative error tolerances, $\epsilon_{abs}$ and $\epsilon_{rel}$, are user defined. In the right hand side of \eqref{eq:errorest1}, the absolute error tolerance is chosen when $\|(v_{xi},v_{yi})_{REF}\|$ is very small and a low relative error may not be necessary. Otherwise, the relative error tolerance is chosen.

\subsubsection{Estimating the error in the residual}\label{sec:Error2}

A computationally cheaper approach is to consider the error in the residual. It measures how well the discretized equations are satisfied by the numerical solution. The residual of the equations in \eqref{eq:momentum} and \eqref{eq:mass} is a right hand side $\res$ which is non-zero if there is an error in the solution. 

Hence, the residual in the numerical solution of \eqref{eq:lineq} should be as small as possible. 

Again let $\tilde{\Delta \uvec}_{SIA}$ be the deviation of $\uvec_{SIA}^{\Omega}$ from the reference solution $\uvec^{\Omega}_{REF}$ in \eqref{eq:Reference}. Then the residual for $\uvec_{SIA}^{\Omega}=\uvec^{\Omega}_{REF}-\tilde{\Delta \uvec}_{SIA}$ can be estimated as
\begin{equation}\label{eq:errorest2}
\tilde{\res}_{SIA} =\b- \tilde{\A}\uvec_{SIA}^{\Omega}=\tilde{\A}(\uvec_{REF}^{\Omega}-\uvec_{SIA}^{\Omega})=-\tilde{\A}\tilde{\Delta\uvec}_{SIA}.
\end{equation}
Only a matrix-vector multiplication is required to compute $\res_{SIA}$. Again, if $\tilde{\A}=\A$, the residual would be exact. The residual corresponding to node $i$ is denoted by $\tilde{\res}_{SIA,i}$. If $\|\tilde{\res}_{SIA,i}\|> \epsilon_{res}$ for a given tolerance $\epsilon_{res}$, then this node is included in $\Omega_{FS}$. The difference between $\uvec_{ISCAL}^{\Omega}$ and $\uvec_{SIA}^{\Omega}$ is non-zero only in $\Omega_{FS}$. Since 
\be\label{eq:FSSSIAdiff}
\begin{array}{rl}
\tilde{\res}_{SIA}^{\Omega_{FS}}&=\b^{\Omega_{FS}}-(\tilde{\A}_{FS}\uvec_{SIA}^{\Omega_{FS}}+\tilde{\A}_{CO}\uvec_{SIA}^{\Omega_{SIA}})\\
                       &=(\tilde{\A}_{FS}\tilde{\uvec}_{FS}^{\Omega_{FS}}+\tilde{\A}_{CO}\uvec_{SIA}^{\Omega_{SIA}})
                         -(\tilde{\A}_{FS}\uvec_{SIA}^{\Omega_{FS}}+\tilde{\A}_{CO}\uvec_{SIA}^{\Omega_{SIA}})\\
                       & =\tilde{\A}_{FS}(\tilde{\uvec}_{FS}^{\Omega_{FS}}-\uvec_{SIA}^{\Omega_{FS}}), 
\end{array}
\ee
the error in the solution and the error in the residual are related as $\tilde{\uvec}_{FS}^{\Omega_{FS}}-\uvec_{SIA}^{\Omega_{FS}}=\tilde{\A}_{FS}^{-1}\tilde{\res}_{SIA}^{\Omega_{FS}}$.

\subsubsection{Estimating the error in a functional of the solution}\label{sec:Error3}
In glaciology, sometimes the velocity field is not of primary interest, but rather some functional of the velocity, such as the ice discharge from a drainage basin (see \cite{Fabien2012}).  Let $f(\uvec)$ be a linear functional of the solution with $f(\mathbf{0})=0$. On the discrete mesh, this functional can be written as $\fatf^T \uvec$, \textit{i.e.} the scalar product of a vector $\fatf$ defining the functional on the mesh and the solution vector $\uvec$. To understand which parts of the ice sheet are important for this functional, the adjoint equation or dual problem
\begin{equation}\label{eq:weight}
\tilde{\A}^T\w=\fatf
\end{equation}
is solved for the weight vector, $\w$, expressing the impact on $f$ of the solution $\uvec$ in each mesh node. We solve the linear system \eqref{eq:weight} with $\tilde{\A}=\A(\uvec_{ISCAL}^{\Omega})$ as in Section \ref{sec:Error1}. Transposing \eqref{eq:weight}, multiplying with the solution error $\Delta \uvec$, and inserting the residual yields the error in the functional $\fatf^T \uvec$ as 
\begin{equation}\label{eq:weight2}
\fatf^T \Delta \uvec = \w^T\tilde{\A} \Delta \uvec =\w^T \res.
\end{equation}
Note that in this case we can either compute $\fatf^T \Delta \uvec$ directly, or evaulate $\w^T \res$. We choose the latter. The residual is evaluated as in Section \ref{sec:Error2}, meaning that $\res=\tilde{\res}_{SIA}$ and $\Delta \uvec = \tilde{\Delta\uvec}_{SIA}$. Let $N$ be the total number of nodes in $\Omega$. If $\w_i$ is the weight vector corresponding to node $i$ and $|\w_i^T\tilde{\res}_{SIA,i}|>\epsilon_f/N$, then the error in the functional due to node $i$ is larger than acceptable based on equidistribution of the error and the user defined error tolerance $\epsilon_f$, and this node is included in $\Omega_{FS}$. If $|\w_i^T\tilde{\res}_{SIA,i}|\le \epsilon_f/N$ for all $i$, then by \eqref{eq:weight2}
\be\label{eq:funcerror}
   |\fatf^T \tilde{\Delta\uvec}_{SIA}| = |\w^T \tilde{\res}_{SIA}|\le \sum_{i=1}^N |\w_i^T\tilde{\res}_{SIA,i}|\le \epsilon_f.
\ee
This way both the model error in SIA measured by the residual $\tilde{\res}_{SIA}$ and the importance of the node $\w_i$ are considered. This approach can be applied to any linear functional $f(\uvec)$. It is the discrete version of the goal oriented or {\it a posteriori} error control in finite elements where the solution to the adjoint differential equation provides the weights $\w$ \cite{EEHJ}. In Section~\ref{sec:Error2}, the nodal weight $\w_i$ is equal to $\tilde{\res}_{SIA,i}$.

\subsection{Computational performance}\label{sec:ComputationalWork}
The computational work for both ISCAL and FS is dominated by solving the Stokes problem. The additional cost of the solution of the free surface problem in \eqref{eq:freesurface}, and other auxiliary processes, can be assumed to be small in comparison. The main computational work when solving any finite element problem is attributed to the assembly of the system matrix and solution of the associated linear system. Often, the solution time dominates the assembly time for large problem sizes, but in some cases the assembly time is significant. For instance in non-linear problems, the solution time is often reduced by using the solution from a previous non-linear iteration as an initial guess for the linear system solver, while the assembly time remains large. When solving the FS system in \eqref{eq:lineq}, the entire system matrix, $\A$, is assembled. In ISCAL, only the smaller matrices $\A_{FS}$ and $\A_{CO}$ are assembled, which is less costly. The solution time is also lower for ISCAL, since \eqref{eq:ISCALsystem} is smaller than the original system \eqref{eq:lineq}.

Let $N_{FS}$ be the number of nodes in $\Omega_{FS}$, and $N_{SIA}=N-N_{FS}$ be the number of nodes in $\Omega_{SIA}$. Assume that the number of non-linear Picard or Newton iterations required to solve \eqref{eq:lineq} and \eqref{eq:ISCALsystem} is $\xi_{FS}$ and $\xi_{ISCAL}$, respectively. Furthermore, assume that the work to solve the linear systems once is $C_{S} N^S$ and $C_S N_{FS}^S$ for FS and ISCAL respectively and that the work for the assembly of the system matrices ($\A$, or $\A_{FS}$) is $C_A N^A$ and $C_A N_{FS}^A$. Both $C_S$ and $C_A$ are constants. The choice of linear solver determines $C_S$ and $S$, and the implementation of the assembly determines $C_A$ and $A$. In ISCAL, there is extra work included in the assembly phase because not only is the system matrix, $\A_{FS}$, assembled but also $\A_{CO}$, and because of the reorganization of \eqref{eq:lineq} into \eqref{eq:ISCALsystem}. We represent this extra load by adding a constant $C_{O}$ to $C_A$. Assuming that any process beside solving the Stokes system is negligible, the speedup, $q$, of ISCAL can be described by,
\be\label{eq:speedup}
\begin{aligned}
q&=\frac{\text{FS load}}{\text{ISCAL load}} \\&= \frac{\xi_{FS}(C_AN^A+C_SN^S)}{\xi_{ISCAL} ((C_A+C_O) N_{FS}^A +C_S N_{
FS}^S +\frac{1}{m\cdot\xi_{ISCAL}}(C_AN^A+C_SN^S))} \, .\\
\end{aligned}
\ee
If the system is solved by Gaussian elimination then $S=3$ and if it is solved by an optimal multigrid algorithm $S$ may approach $1$ \cite{BSA14}. The system matrix is sparse, and so $A=1$.

The term $(C_AN^A+C_SN^S)/(m \xi_{ISCAL}$) represents the error estimation. It is small because of the factor $1/(m \xi_{ISCAL})$, \textit{i.e.} because the error estimation is not done in every non-linear iteration, and because the error estimation is typically not done in every timestep, \textit{i.e.} $m>1$. If the residual based error estimate is used, $C_SN^S$ is removed.

If we assume that $\xi_{ISCAL}$ is almost equal to $\xi_{FS}$, and that $C_O$ is not much greater than $C_A$, then we have $q>1$, since $N_{FS}<N$. In Section \ref{Efficiency}, the computational gain of using ISCAL is measured in numerical experiments and \eqref{eq:speedup} is revisited.

\section{Numerical Experiments} \label{sec:numexp}

In this section four experiments are performed. In Experiment 1, the ISCAL is evaluated for accuracy and efficiency. The ability to dynamically change the FS domain $\Omega_{FS}$ is tested in Experiment 2. In Experiment 3, the three different error estimates are compared. ISCAL is tested on real data in Experiment 4, . In Experiments 1-3 a conceptual model problem (see Section \ref{sec:modeldescrip}) is considered, and in Experiment 4 data from the Greenland Ice Sheet is used. The simulations are performed on a single core, although a parallel version of ISCAL now exists and is used in \cite{QSR}.

\subsection{Model problem description}\label{sec:modeldescrip}
The geometry of the first three experiments is a 3D circular ice sheet. The upper surface position, $h(x,y,t)$, is initialized as a so-called Vialov profile \cite{GreveBlatterBok}
\begin{equation}
h(x,y,0)=h_0\left(1-\left(\frac{r}{L}\right)^{(n+1)/n} \right)^{n/(2n+2)}+100,
\end{equation}
and evolves in time according to \eqref{eq:freesurface}. Here $h_0=3575.1$ m is the maximum initial height, $L=750$ km is the radius of the ice sheet, and $r$ is the distance of a point $(x,y)$ from the center. By comparison, the Greenland Ice Sheet is about $2400$ km long and $1100$ km wide, and is about $2000$ to $3000$ m thick.

The bedrock is flat (\textit{i.e.} $b=0$), and since the problem is isothermal, the rate factor is constant, $\mathcal{A}=10^{-16}\, {\rm Pa}^{-3}\, {\rm year}^{-1}$. The Glen parameter, $n$, is set to the standard value $3$, the density, $\rho$, is $910\, {\rm kg}\, {\rm m}^{-3}$, and the accumulation/ablation function is given by
\begin{equation}
a_s = \min\{0.5,10^{-5}(450\cdot 10^3-\sqrt{x^2+y^2}) \}
\end{equation}
m/year, so that the the accumulation is positive in the center of the ice sheet and decreases radially, becoming negative at $r=450$ km, mimicking accumulation/ablation patterns observed in natural ice sheets. Note that this is not the accumulation leading to the Vialov profile in \cite{GreveBlatterBok}, but rather the accumulation function used in the EISMINT benchmark experiment \cite{Huybrechts}. At the base of the ice, a no-slip condition is applied for Experiment 1 and 3, and a linear slip condition for Experiment 2. 

The mesh is constructed by vertically extruding a 2D footprint mesh to 20 layers (see \fgref{fig:ExtrudedMesh}). The 2D footprint mesh is unstructured and triangular in the center of the circle, and radially aligned towards the margins.

In all experiments, the linear system for the Stokes problem is solved iteratively with the generalized conjugate residual method, and the non-linear equation is solved with a Picard iteration. An under-relaxation with factor 0.9 is applied to the Picard iteration in order to stabilize it.

\subsection{Experiment 1 - Efficiency and accuracy}\label{Efficiency}

\subsubsection{Setup}
The purpose of ISCAL is to reduce the simulation time with a controlled, small reduction in accuracy. In this experiment, four different meshes of varying resolution are employed to simulate the model problem with ISCAL, FS, and SIA  in order to compare simulation time and accuracy. The coarsest mesh has $17860$ nodes (see \fgref{fig:ExtrudedMesh}), and the finest mesh has $257000$ nodes. The number of nodes in the finest mesh is of the same order of magnitude as the number of nodes used for simulating the Greenland Ice Sheet in \cite{Hakime2012} and \cite{Fabien2012}. Since the number of non-linear iterations varies in time, we run the simulation for $30$ time-steps of one month each. When measuring the CPU-time, the un-representative first time-step is excluded. In ISCAL, we estimate the error in the horizontal velocities every tenth time step as in \eqref{eq:errorest1}, with the relative tolerance, $\epsilon_{rel}$, set to $5\%$ and the absolute tolerance $\epsilon_{abs}=1$ m/year. 

\subsubsection{Results}

\begin{figure}
    \centering
   {{\includegraphics[trim={5cm 13cm 1cm 5cm},clip,width=0.5\textwidth]{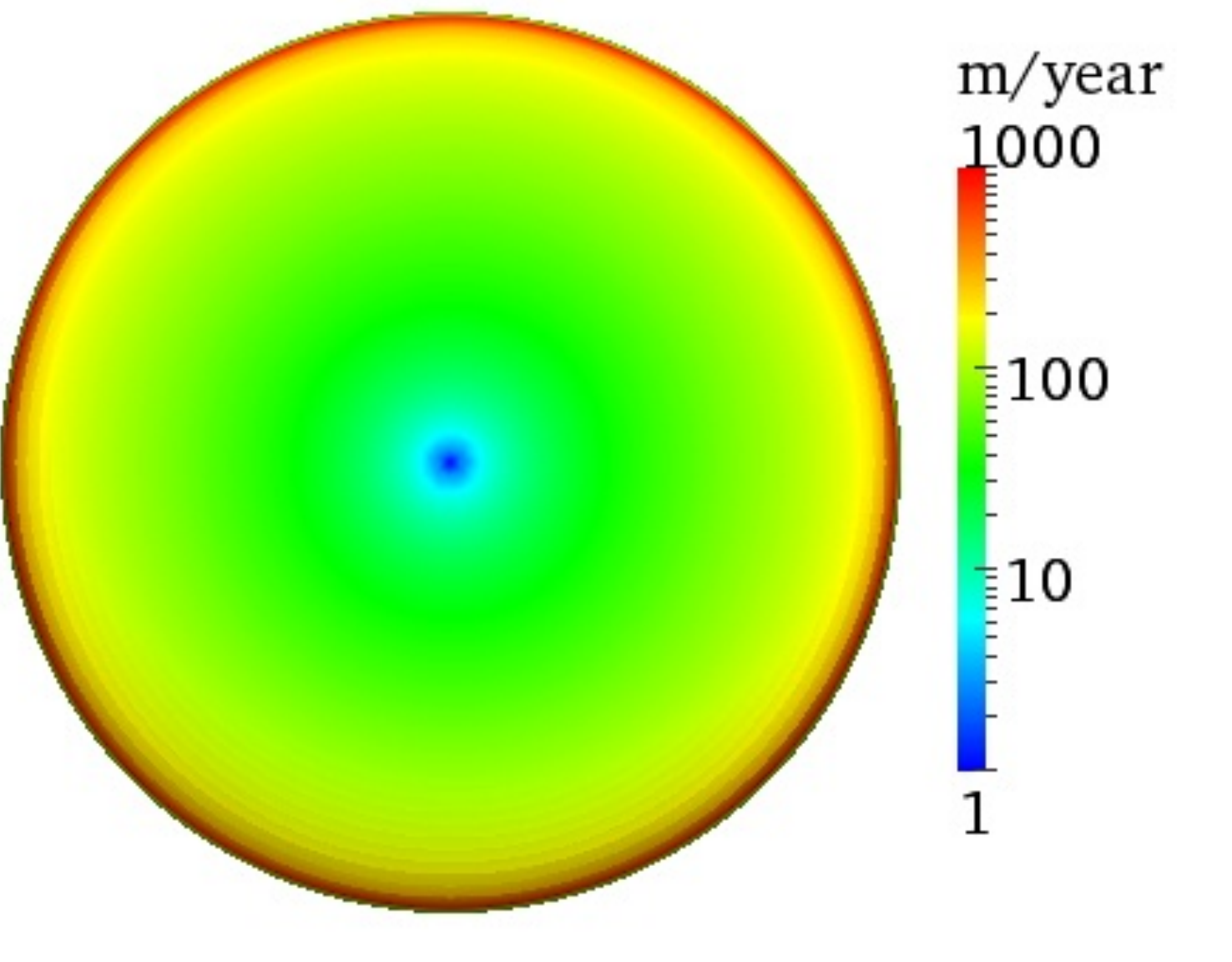} }}%
    \caption{Experiment 1 - The velocity magnitude of a circular ice sheet, as seen from above, after $30$ months. The solution is computed with ISCAL on a mesh with $96520$ nodes.}%
    \label{fig:ISCALVelocity}%
\end{figure}

\begin{figure}
    \centering
    \subfloat[Relative Error in SIA, $\Delta\uvec_{SIA}$]{{\includegraphics[trim={5cm 18cm 1cm 1cm},clip,width=0.45\textwidth]{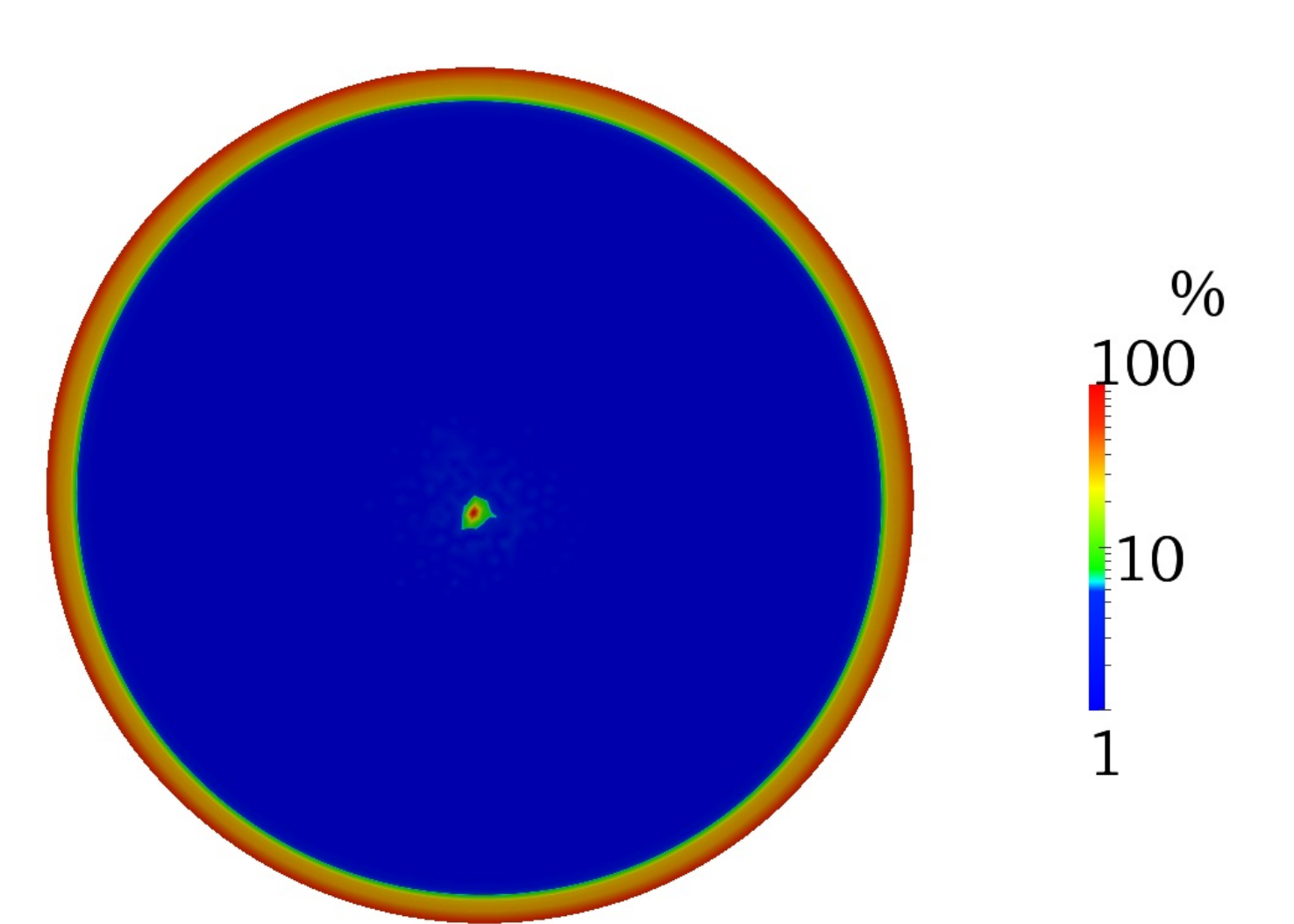} }  \label{fig:Errors1a}}%
    \qquad
    \subfloat[Estimated Error, $\tilde{\Delta\uvec}_{SIA}$]{{\includegraphics[trim={5cm 17.7cm 1cm 1cm},clip,width=0.44\textwidth]{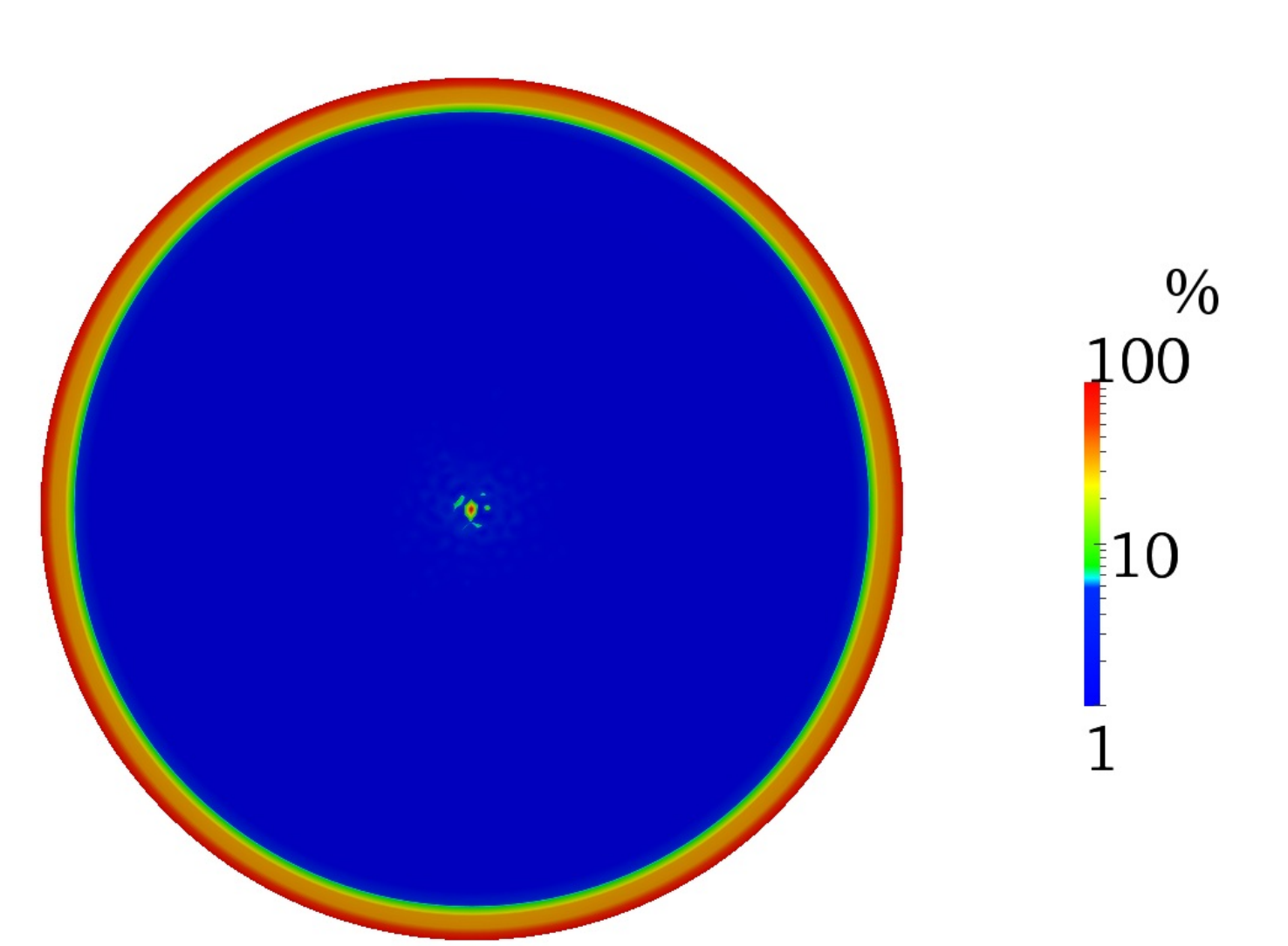} }\label{fig:Errors1b}}%
    \qquad
    \caption{Experiment 1 - Exact and estimated relative error in SIA in percent, after $30$ months, viewed from above. The relative error tolerance, $\epsilon_{rel}$, in SIA is 5 $\%$ and the absolute error tolerance, $\epsilon_{abs}$ is $1$ m/year. The color scale is set so that everything less than $5$ \% is blue.}%
\label{fig:Errors1}
\end{figure}

\begin{figure}
    \centering
    \subfloat[$\Omega_{FS}$ (red) and $\Omega_{SIA}$ (blue)]{{\includegraphics[trim={5cm 17.7cm 1cm 1cm},clip,width=0.44\textwidth]{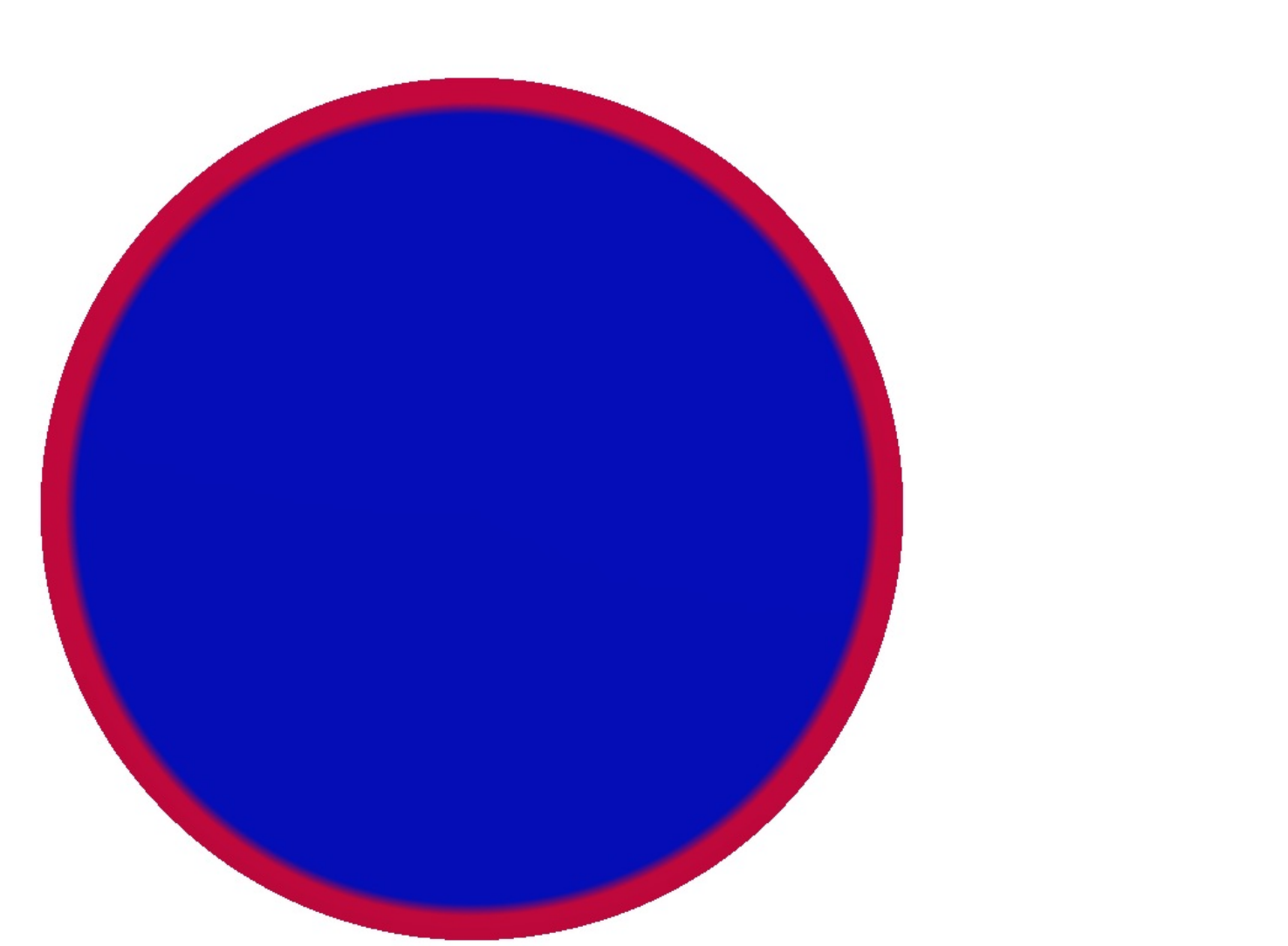} }\label{fig:Omegasa}}%
\qquad
 \subfloat[Relative Error in ISCAL, $\tilde{\Delta\uvec}_{ISCAL}$]{{\includegraphics[trim={5cm 18cm 1.1cm 1.1cm},clip,width=0.45\textwidth]{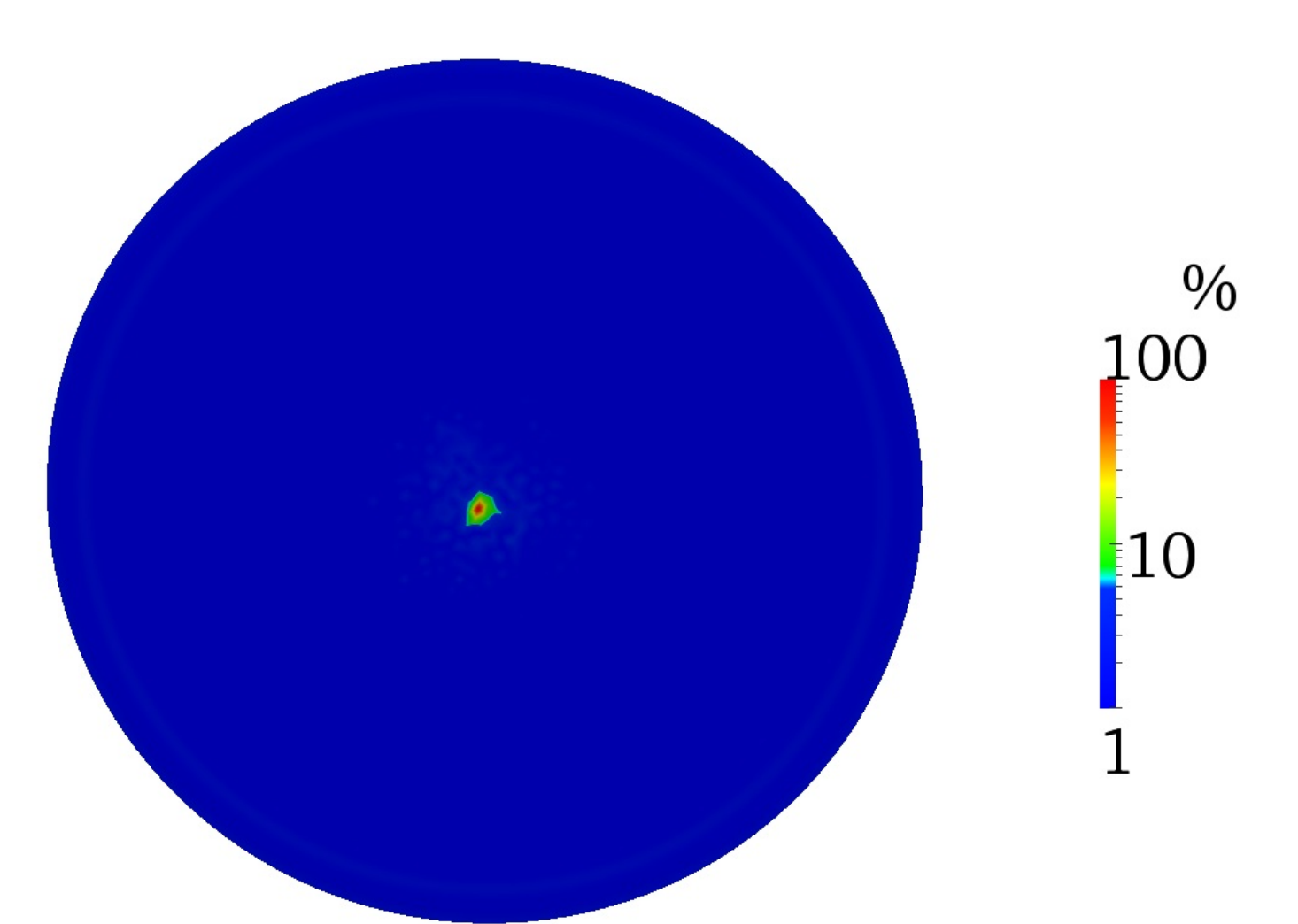} }\label{fig:Omegasb}}%
    \qquad
    \caption{Experiment 1 - The distribution of $\Omega_{FS}$ (red) and $\Omega_{SIA}$ (blue) and the error in ISCAL compared to FS after $30$ time-steps. }%
\label{fig:Omegas}
\end{figure}

The magnitude of the velocity is high at steep margins (1000 m/year or more), and low near the dome, (\fgref{fig:ISCALVelocity}). The same pattern is observed in nature, \textit{e.g.} in the Greenland Ice Sheet \cite{Ian}. The velocity field in \fgref{fig:ISCALVelocity} is computed by ISCAL. The FS and SIA solutions are not shown since the difference between ISCAL, FS, and SIA velocity is difficult to discern by the naked eye (using a reasonable color scale), despite considerable relative errors in the SIA (\fgref{fig:Errors1a}).
As expected from theory (\textit{e.g.} \cite{BlaHaftet,Hutter83,Ahlkrona13a}), the SIA error is particularly high near the margins and at the dome (in some places greater than $100$ \% ) (\fgref{fig:Errors1a}). ISCAL automatically detects these problematic regions through the error estimation, and applies the FS where the error is higher than $5$ \% or $1$ m/year (\fgref{fig:Omegasa}). The estimated error from \eqref{eq:estimatederror}, shown in \fgref{fig:Errors1b}, is very close to the true error from \eqref{eq:trueerror} and hence is a good base for the error control. Note that the linkage between the SIA and the FS is very smooth, and not visible in \fgref{fig:ISCALVelocity}.
The relative error of ISCAL compared to the solution of the FS equations is much lower than for the SIA (\fgref{fig:Omegasb}). In this case, the FS is not applied at the dome even if the relative error is high there. This is so because the horizontal velocity is very low at the dome and we allow for an absolute horizontal velocity error of $1$ m/year, see \eqref{eq:errorest1}. 

\begin{figure}
    \centering
    {{\includegraphics[width=0.7\textwidth]{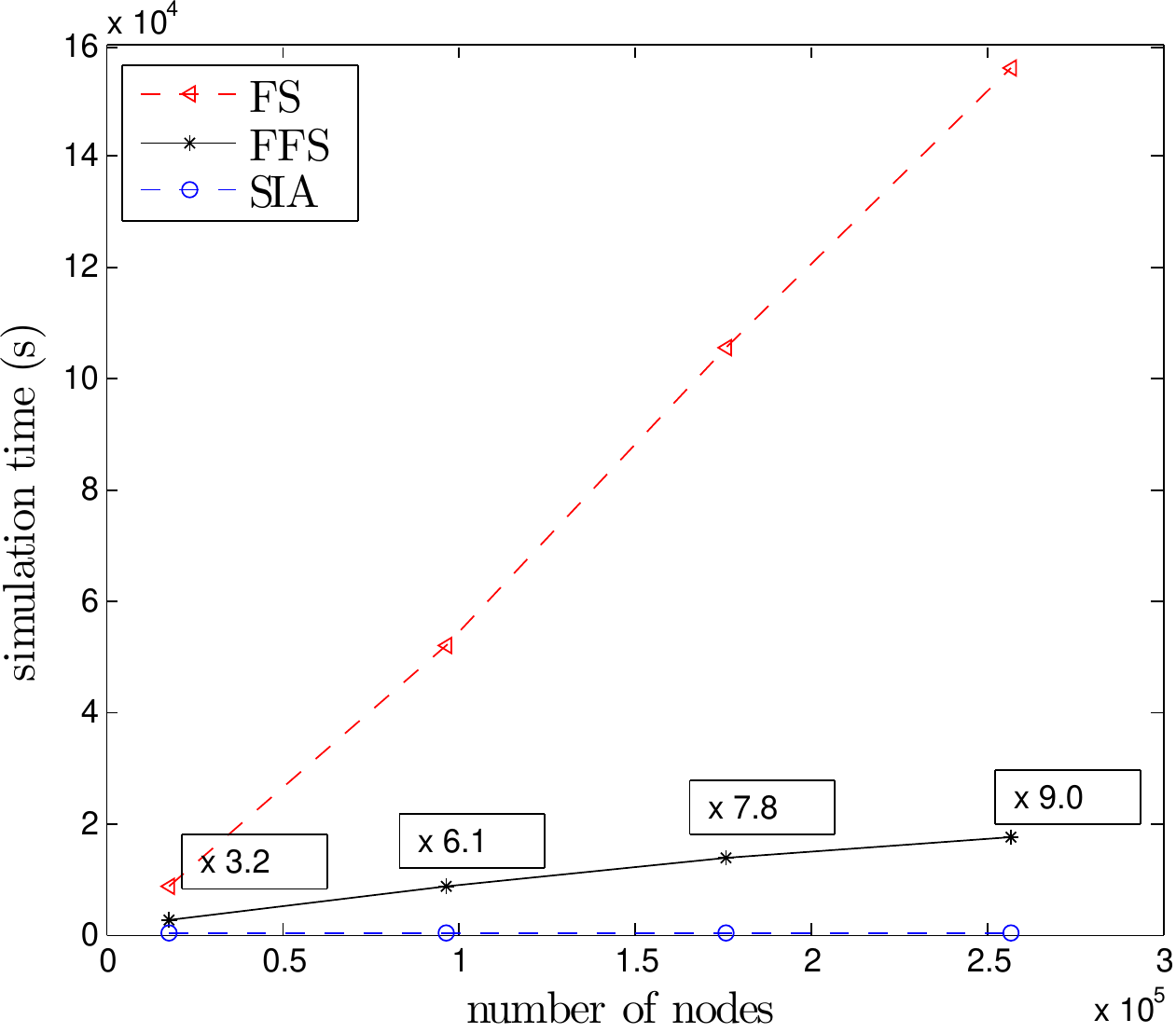} }}%
    \caption{Experiment 1 - CPU-time versus number of nodes for the SIA (blue dashed line), ISCAL (black solid line) and the full Stokes (red dashed line) problem. The labels indicate speedup, $q$, of ISCAL relative to FS.}%
    \label{fig:cputime}%
\end{figure}

\begin{figure}
    \centering
    {{\includegraphics[width=0.7\textwidth]{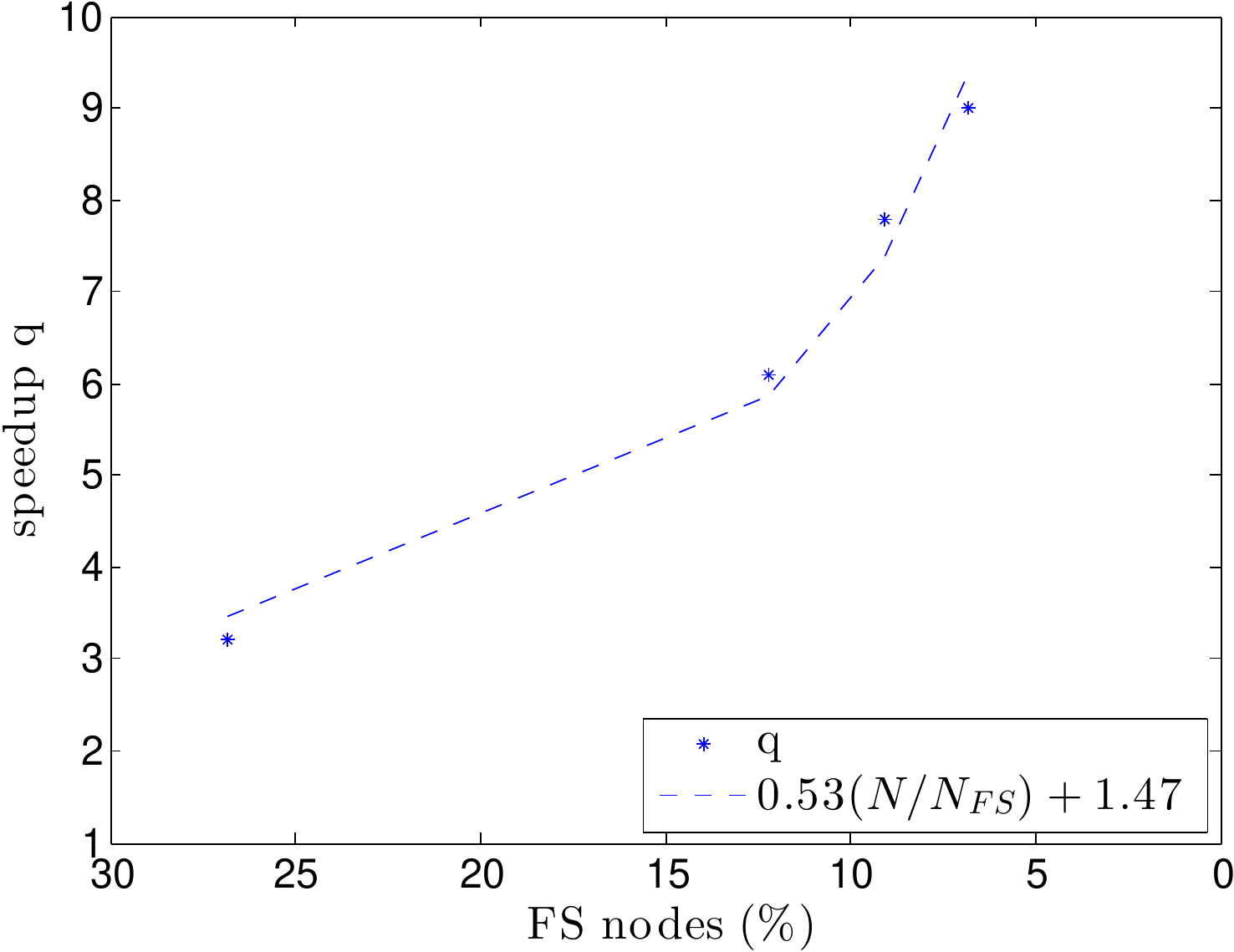} }}%
    \caption{Experiment 1 - Speedup $q$ versus percentage of FS nodes (stars). The dashed lines indicates the polynomial fit $q=0.53(N/N_{FS})+1.47$. In our experiments, the larger the problem size, the smaller the percentage of FS nodes.}%
    \label{fig:speedup}%
\end{figure}

The CPU-time of the ISCAL is considerably lower than for the FS system (\fgref{fig:cputime}). The speedup increases with problem size, and on the finest mesh ISCAL is nine times faster than the FS. The CPU-time required to compute the SIA is negligible compared to both FS and ISCAL, which is why it is so popular.
\tbref{tab:times} provides more insight in what operations are the most time-consuming, for both FS and ISCAL, on all four meshes.
\begin{table}[h]
    \caption{Experiment 1 - Percentage of full Stokes nodes, percentage of CPU-time spent on matrix assembly, linear system solution, and error calculation, and average number of non-linear iterations per time step for the four different meshes.}
\begin{tabular}{ l l || c | c | c | c | c }
  \# nodes, $N$ & Model & FS nodes &Assembly &  Solve & Error Calc. &  \# iter.  \\
  & &  (\%)& (\%)  & (\%)   & (\%) & \\
\hline          \hline
 \multirow{2}{*}{ $17860$} & FS & 100.0 &87.0 & 12.3& - & 10.4   \\ 
  & ISCAL &26.8 &84.8 & 10.7& 2.1 & 10.2 \\ \hline
 \multirow{2}{*}{  $96520$} & FS & 100.0&  86.6& 12.7& -&  11.6 \\
  &ISCAL & 12.2&  82.3& 9.6 & 3.8 &   11.9 \\ \hline
  \multirow{2}{*}{   $175960$ }& FS & 100.0&  85.7& 13.7& -& 12.7\\
  &ISCAL &9.1 &  82.0&  8.8& 4.4 &   12.7 \\ \hline
  \multirow{2}{*}{   $257000$ }&  FS& 100.0& 84.8 & 14.6 &- & 12.9 \\
  & ISCAL&  6.8&  78.3&  11.2&  5.1&   12.9 \\ 
  \hline  
\end{tabular}
\label{tab:times}
\end{table}
For FS, the assembly and solution of the Stokes system completely dominates other processes, such as calculation of the SIA or solving for the free surface position, as assumed in \eqref{eq:speedup}. For ISCAL, also the error estimation time is significant, although still much smaller than the time for the assembly and solution. The relative number of FS nodes decreases with problem size, most likely because the numerical errors in both SIA and FS are reduced with increasing resolution. Because of this, the relative time spent on the error estimation increases with the problem size, whereas the assembly and solution time decreases, while the error estimation time is independent from the size of $\Omega_{FS}$. The assembly time dominates the solution time, even though the importance of the solution time increases slightly with problem size as expected if $S>1$. The average number of non-linear iterations per time step increases slightly with problem size, and is the same for ISCAL and FS, because the areas that converge slowly are included in $\Omega_{FS}$. By revisiting \eqref{eq:speedup}, using the data in \tbref{tab:times}, we can better understand how the speedup, $q$, is related to $N/N_{FS}$. If we use that 1) $C_A>C_S$ in \eqref{eq:speedup}, 2) $S$ is not very much larger than unity, 3) the time spent on error estimation is small, and 4) that $\xi_{FS}\approx \xi_{ISCAL}$, we can roughly approximate \eqref{eq:speedup} by
\be\label{eq:speedupapprox}
q = \frac{C_A}{C_A+C_O} \frac{N}{N_{FS}}+\text{small remainder}.\\
\ee
Fitting $q(N/N_{FS})$ to a linear polynomial gives $C_A/(C_A+C_O)\approx 0.53$, such that the speedup is less than $N/N_{FS}$ (\fgref{fig:speedup}). 
In other words, the computational work associated with rearranging \eqref{eq:lineq} to \eqref{eq:ISCALsystem}, and the assembly of $\A_{CO}$, is of the same order of magnitude as the assembly of $\A_{FS}$ in the present implementation. This reduces the speedup, $q$. In a general context, often $C_A<C_S$, which would make the speedup larger. The speedup would also be larger if a direct solver is used, since then $S \approx 3$. As indicated in \fgref{fig:speedup}, increased speedup with problem size can be attributed to the fact that the proportion of FS nodes decreases with increasing problem size. There are, however, additional explanations, neglected in \eqref{eq:speedupapprox}, \textit{e.g.} that the overhead from sorting the nodes has a larger relative impact for small problem sizes, and that $S>1$.

\subsection{Experiment 2 - Ice stream tracking}

\subsubsection{Setup}
In this experiment, we test how well ISCAL adapts to rapid changes in dynamics, and how it treats fast flowing ice streams. Ice streams are very important for the mass budget of the Greenland Ice Sheet, and even more so for the Antarctic Ice Sheet \cite{Rignot,JoughinAnt}. As the SIA assumes that shearing motion dominates sliding, we expect ISCAL to apply the Stokes equations in ice streams. Ice streams may change position during long time spans, 
\cite{Dowdeswell,Winsborrow2,Conway} and $\Omega_{FS}$ should adapt to this change. In order to test this, the basal boundary condition is changed from no slip to the linear sliding law in \eqref{eq:slidinglaw} with 
\be
\beta(x,y,t)=\max\left(10^{-4},10^{\left(-3.0e^{\frac{(\theta(x,y)-t)^2}{0.18}}-0.7\right)}\right),
\ee
where $\theta(x,y)$ is the polar angle of every point $(x,y)$. This sliding law allows for an ice stream to form, starting at the center and flowing radially outwards to the margin, moving counter clockwise as the time $t$ increases. The area outside the ice stream has a very high friction ($\beta=10^{-0.7}$) effectively imposing a no slip condition. ISCAL is applied using the error estimation based on horizontal velocity (Section~\ref{sec:Error1}) with a relative tolerance, $\epsilon_{rel}$, of $5$ \% and $\epsilon_{abs}=1$ m/year, on a mesh with $96520$ nodes. The error is estimated in every time-step. If the error estimation is performed less frequently, there will be a delay in the update of $\Omega_{FS}$ and $\Omega_{SIA}$.

\subsubsection{Results}

\begin{figure}
    \centering
    {{\includegraphics[trim={5cm 13cm 1cm 5cm},clip,width=0.5\textwidth]{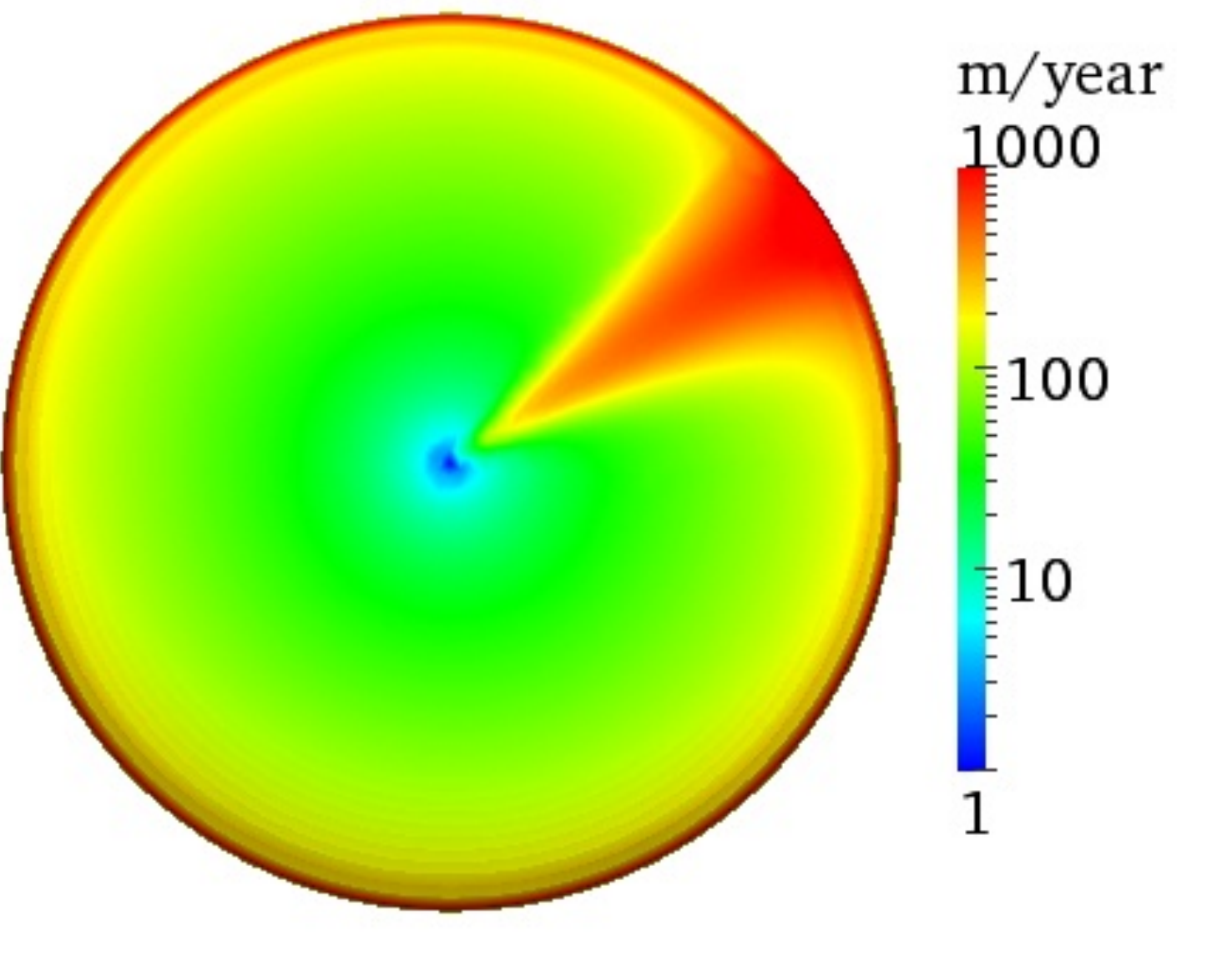} }}\\%
    \caption{Experiment 2 - Velocity magnitude with ISCAL after 11 months, seen from above. The boundary between $\Omega_{FS}$ and $\Omega_{SIA}$ is not visible.}%
\label{fig:Tracking2}
\end{figure}

\begin{figure}
    \centering
    \subfloat[$1$ month]{{\includegraphics[trim={5cm 13cm 5cm 1cm},clip,width=0.26\textwidth]{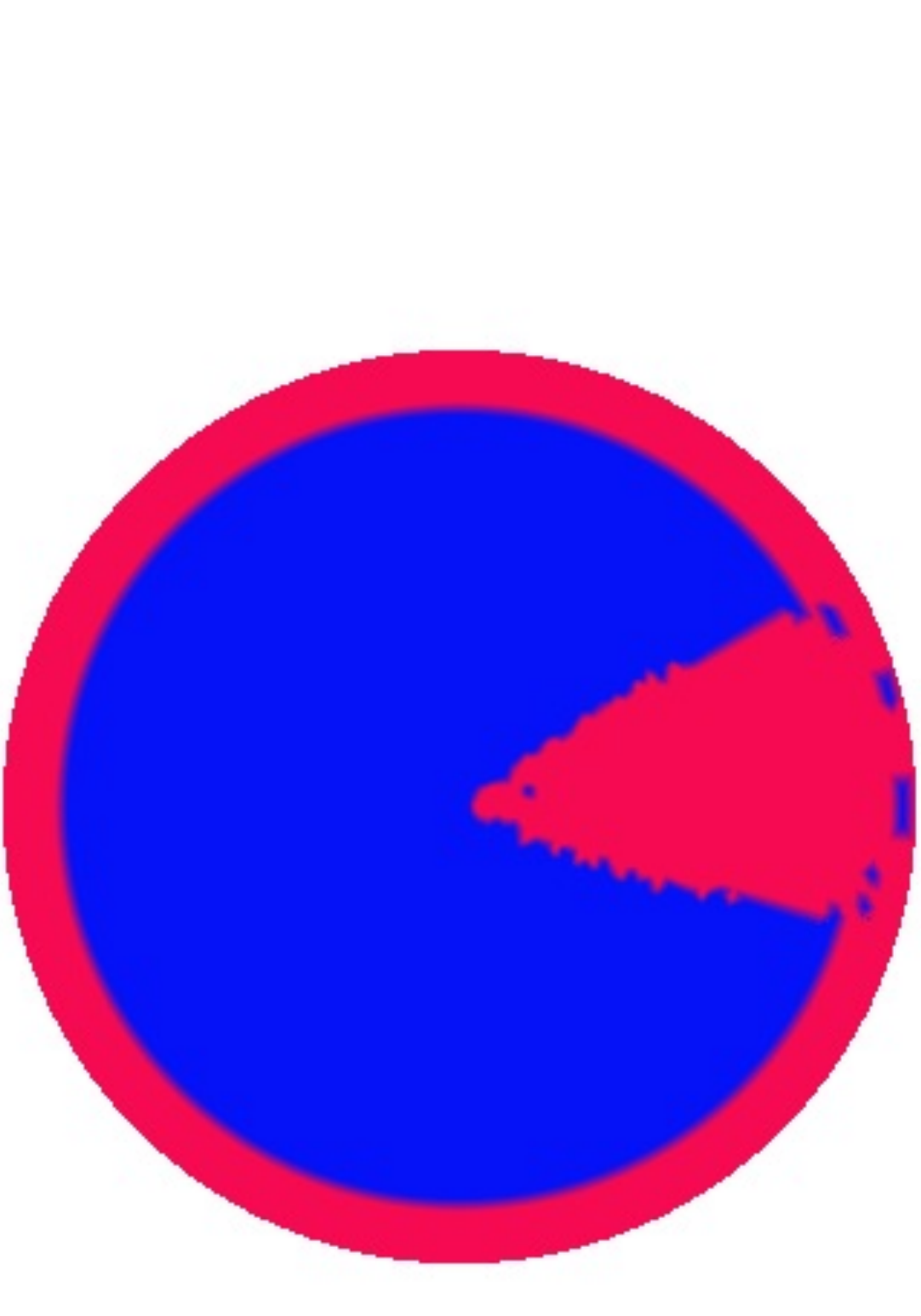} }}%
    \qquad
    \subfloat[$6$ months]{{\includegraphics[trim={5cm 13cm 5cm 1cm},clip,width=0.26\textwidth]{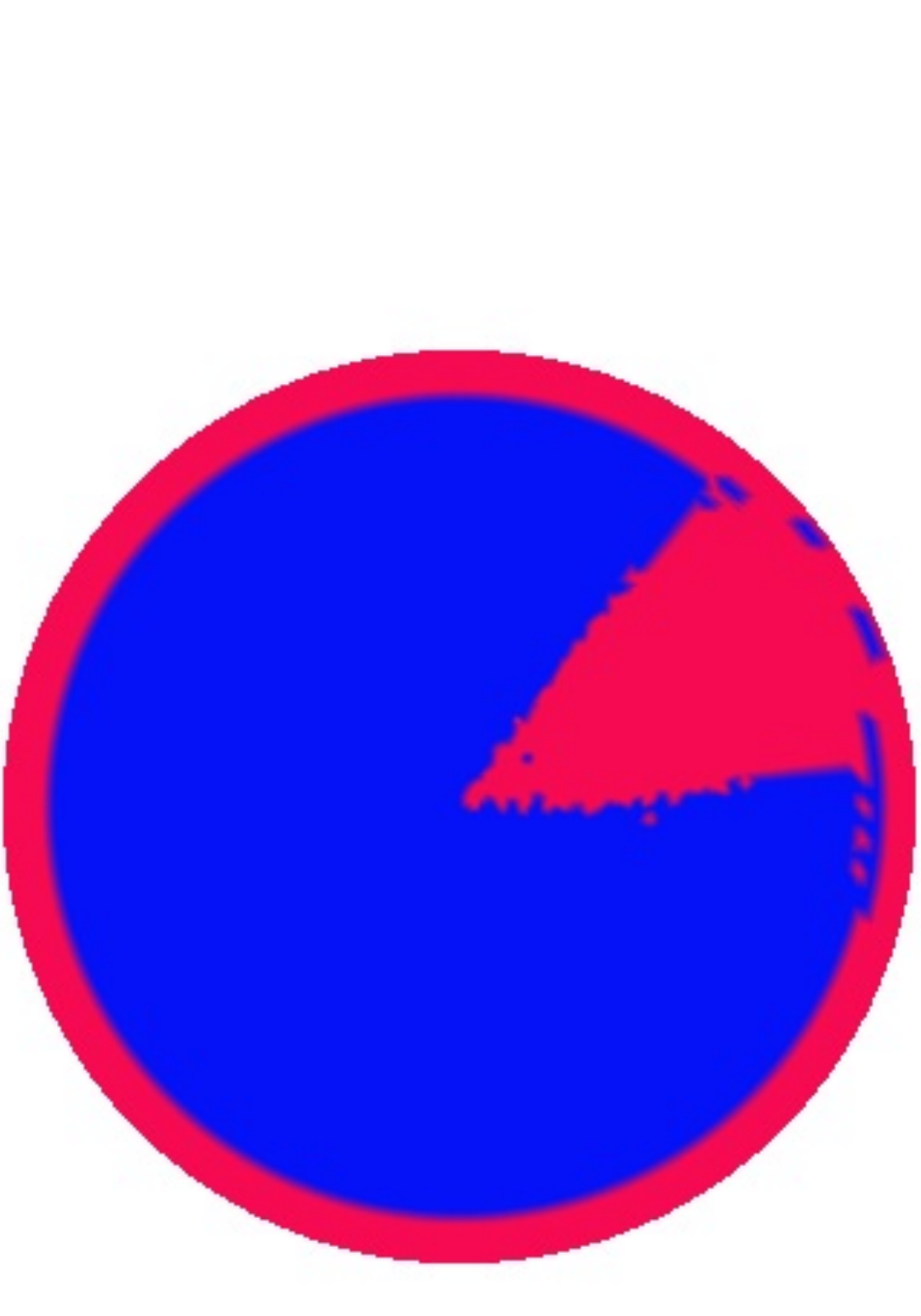} }}%
    \qquad
  \subfloat[$11$ months]{{\includegraphics[trim={5cm 13cm 5cm 1cm},clip,width=0.26\textwidth]{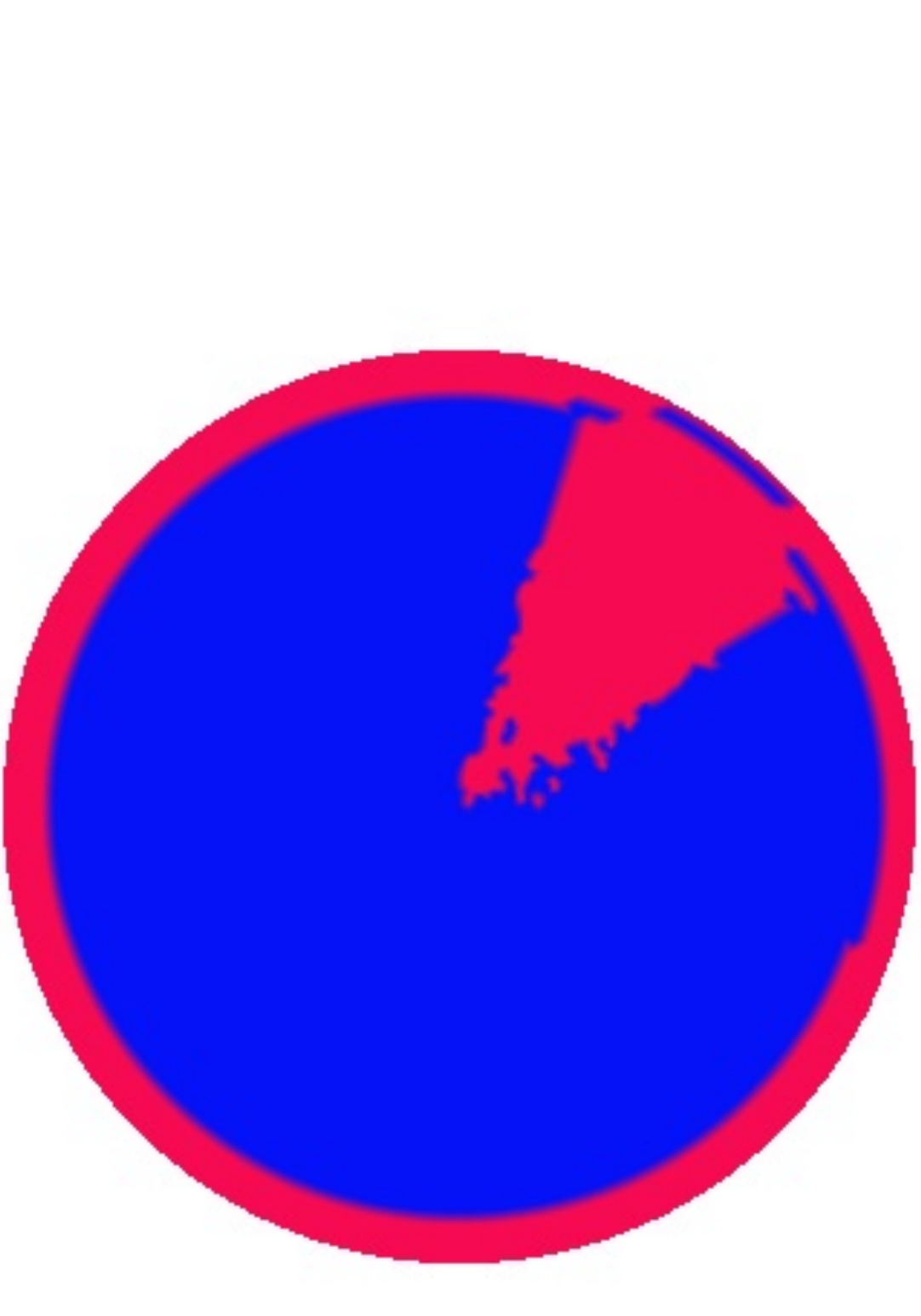} }}%
    \caption{Experiment 2 - Distribution of $\Omega_{FS}$ (red) and $\Omega_{SIA}$ (blue) after $1$, $6$ and $11$ months. The ice stream position switches counter clockwise.}%
\label{fig:Tracking}
\end{figure}

The modified sliding law creates an ice stream in which the velocity is significantly higher than in the surrounding ice (\fgref{fig:Tracking2}).
ISCAL automatically tracks this ice stream as it moves, applying the Stokes equations in the fast flowing region and at the margins as expected (\fgref{fig:Tracking}). 
\subsection{Experiment 3 - Comparing error estimations}

\subsubsection{Setup}

The three different methods of error estimation presented in Section~\ref{sec:ErrorEstimation} will lead to different distributions of the FS nodes, \textit{i.e.} of $\Omega_{FS}$ and $\Omega_{SIA}$. To compare the different distributions we run three simulations using each one of the three estimates. The set-up is the same as in Experiment 1, and the simulations run for $12$ time-steps (months) using the mesh with $96520$ nodes. We keep the number of nodes in $\Omega_{FS}$ constant at $20$ \% of the total (\textit{i.e.} $19304$ nodes), such that the $19304$ nodes with the highest error, computed with respective error measures, are included in $\Omega_{FS}$. When estimating the error in the velocity field, the nodes are sorted based on the relative error in the velocity, but a node that has an absolute error that is lower than an absolute tolerance of $0.1$ m/year will not be included in $\Omega_{FS}$. This prevents areas with almost zero velocity (\textit{e.g.} near the base) from inclusion in $\Omega_{FS}$. The absolute tolerance is was lowered compared to the previous experiments, in order to avoid repetition. For the functional based error estimation, (Section~\ref{sec:Error3}), the functional is chosen as the ice flux across the surface $F(x,y):\sqrt{x^2+y^2}=600$ km 
\be
f(\uvec)=\int_F \mathbf{n_f}^T \v\,dA,
\ee
where $\n_f$ is the normal pointing outwards from the surface $F$. In a real ice sheet, a large part of the mass loss is through ocean terminating ice streams. A functional describing the ice flux across a vertical surface can be used to control the error in the discharge through such streams. The simulation where the error in the residual is estimated is straightforward.

\subsubsection{Results}

\begin{figure}
    \centering
    \subfloat[Horizontal Velocity]{{\includegraphics[trim={5cm 20cm 5cm 4cm},clip,width=0.38\textwidth]{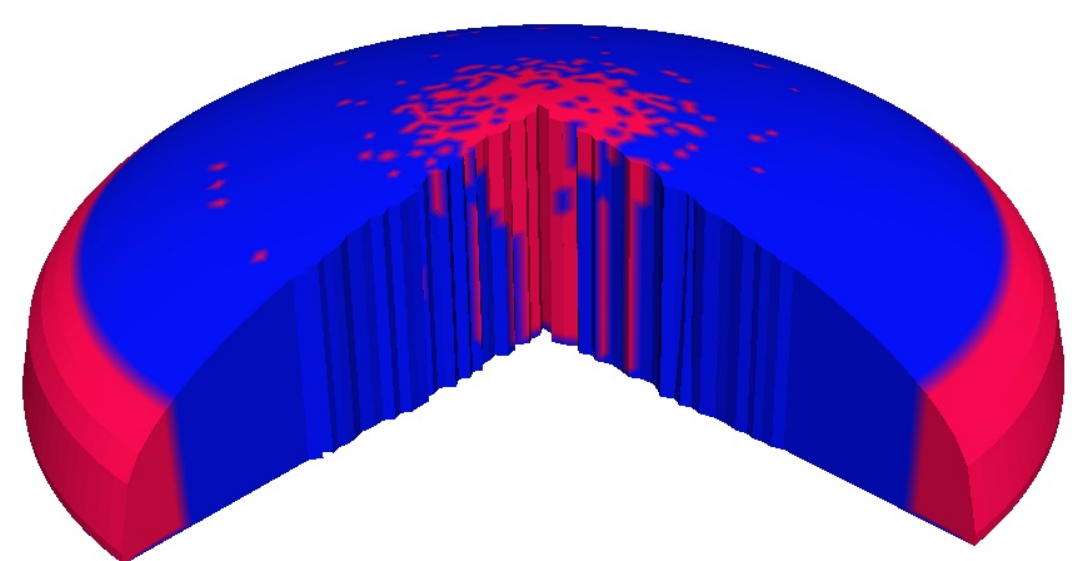} }\label{fig:Distributiona}}%
    \qquad
    \subfloat[Flux]{{\includegraphics[trim={5cm 20cm 5cm 4cm},clip,width=0.38\textwidth]{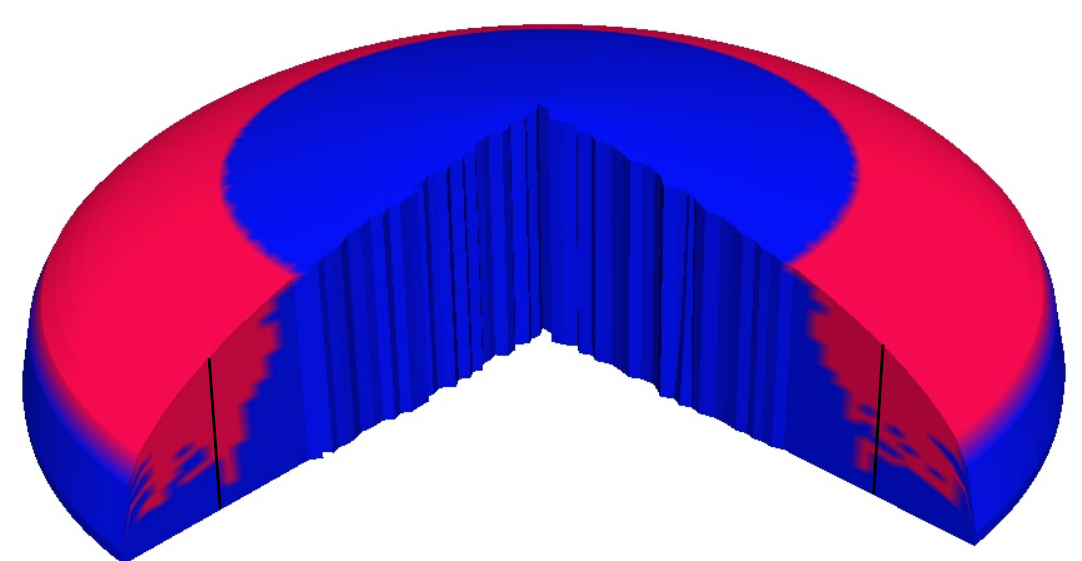} }\label{fig:Distributionb}}%
    \qquad
  \subfloat[Residual]{{\includegraphics[trim={5cm 20cm 5cm 4cm},clip,width=0.38\textwidth]{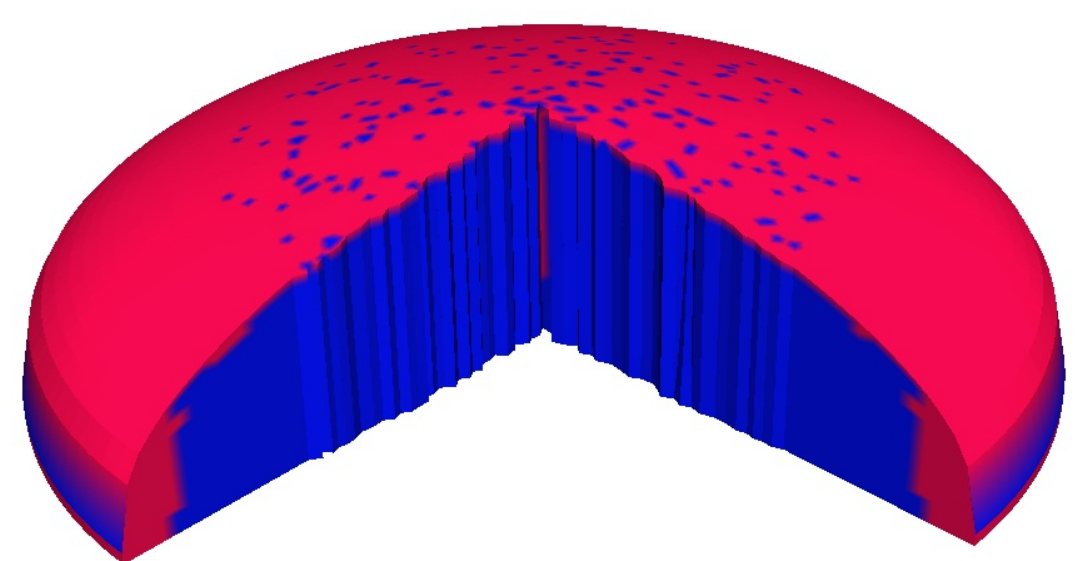} }\label{fig:Distributionc}}%
    \caption{Experiment 3 - Distribution of $\Omega_{FS}$ (red) and $\Omega_{SIA}$ (blue) for the three different error estimations after $12$ months. A slice is cut out to show how the FS nodes are distributed within the ice, and the vertical axis is scaled by a factor 100. The vertical black lines in (b) indicate the position of the surface $F:\sqrt{x^2+y^2}=600$ km.}%
\label{fig:Distribution}
\end{figure}

As expected, the FS area, $\Omega_{FS}$, depends on the method of error estimation (\fgref{fig:Distribution}).
When estimating the error in the horizontal velocity, the distribution is similar to the previous experiments, \textit{i.e.} FS is applied at the margins (\fgref{fig:Distributiona}), but as the absolute tolerance was lowered from $1$ m/year to $0.1$ m/year also the dome is now included in $\Omega_{FS}$. When estimating the error in the ice flux, the FS nodes are concentrated around the surface $F(x,y)$ (\fgref{fig:Distributionb}). As there is no sliding at the base, the main motion is by vertical shearing, such that the flux through the surface is higher near the surface than close to the base. Therefore there will be more FS nodes near the surface than near the base, as observed in \fgref{fig:Distributionb}. The simulation estimating the error in the residual results in FS nodes at the dome, near the margins, and close to the ice surface (\fgref{fig:Distributionc}). Since the solution is numerically sensitive near the surface, the distribution of nodes is somewhat irregular there. The same effect is found in \fgref{fig:Distributiona}. The magnitude of the error is higher at the margins than at the dome and at the surface. Thus if the number of FS nodes is decreased, $\Omega_{FS}$ is similar for the residual based and the velocity based error estimates. The computational work to estimate the error in the residual is lower than in the other error estimates, because no linear system is solved. On the other hand it is perhaps not as intuitive as estimating the error in the velocity or in the flux. 

\subsection{Experiment 4 - Application to the Greenland Ice Sheet} \label{sec:Greenland}

\subsubsection{Setup}

In order to test ISCAL on realistic data of interest to the glaciological community, we apply the ISCAL to the Greenland Ice Sheet. Note that we are not aiming to obtain results that give insight into the ice sheet dynamics of Greenland nor to answer any glaciological questions. We would like to emphasize that we simply want to use the Greenland data in order to test the method on a real-world geometry, rather than to present a complete simulation of the Greenland Ice Sheet. The application of the method to problems of glaciological interest will be reported in future studies. For simplicity, we consider isothermal conditions (with $\mathcal{A}$ as in previous experiments), do not apply any time-integration and thus need no climatic forcing (\textit{i.e.} $a_s=0$). The data required are bedrock topography, $b$, under the ice sheet, ice surface topography, $h$, and a sliding coefficient $\beta$ describing the friction between the ice and the bedrock. The sliding coefficient is incorporated in the linear sliding law of \eqref{eq:slidinglaw}.

\begin{figure}
    \centering
    \subfloat[Surface topography, $h$]{{\includegraphics[trim={6cm 10cm 5.8cm 0.9cm},clip,width=0.27\textwidth]{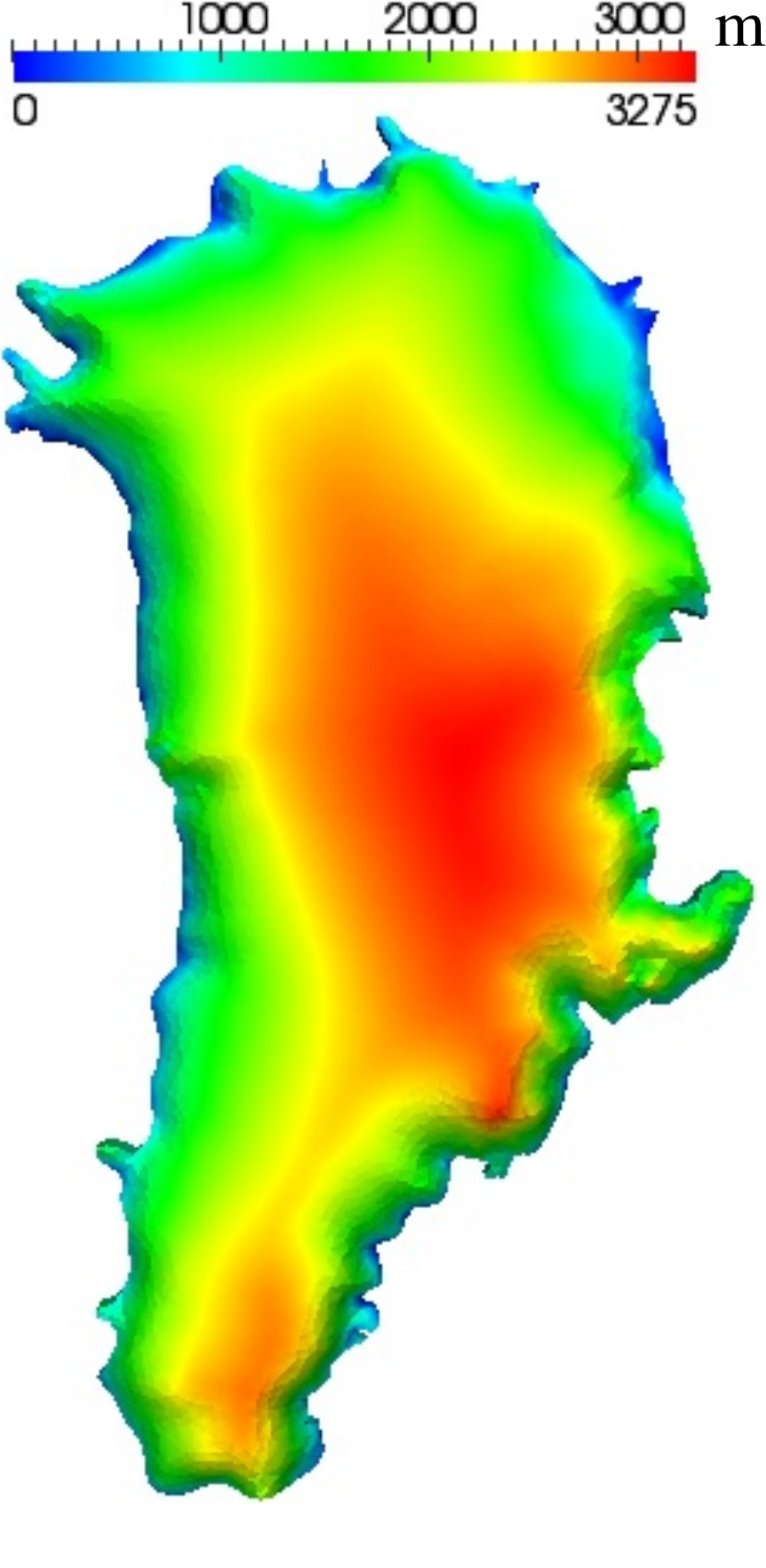} }\label{fig:geometrys}}%
    \qquad
    \subfloat[Basal topography, $b$]{{\includegraphics[trim={6cm 10cm 5.5cm 1cm},clip,width=0.28\textwidth]{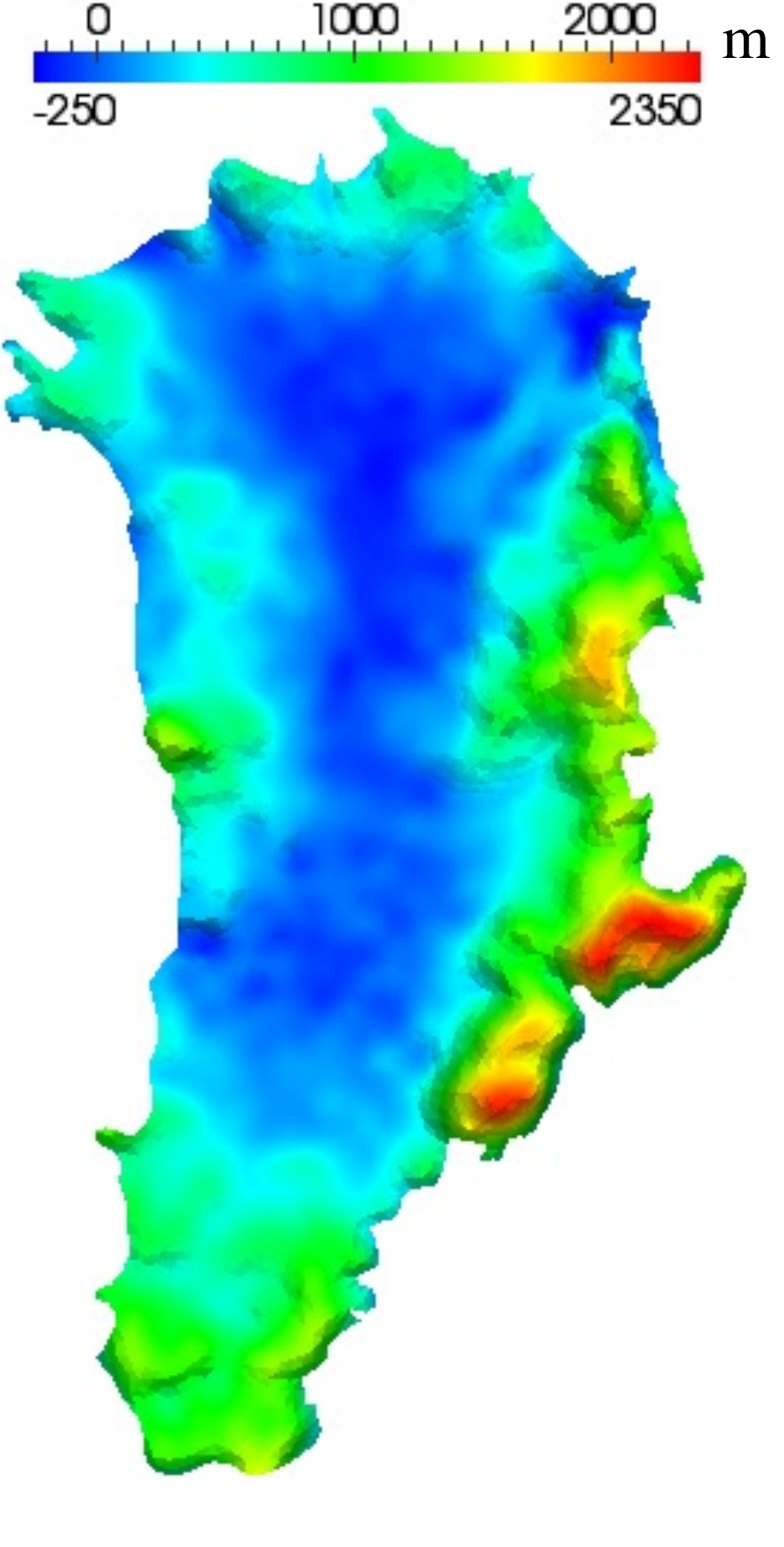} }\label{fig:geometryb}}%
    \qquad
    \subfloat[Basal sliding coefficient, $\beta$]{{\includegraphics[trim={6cm 10cm 5.8cm 1cm},clip,width=0.27\textwidth]{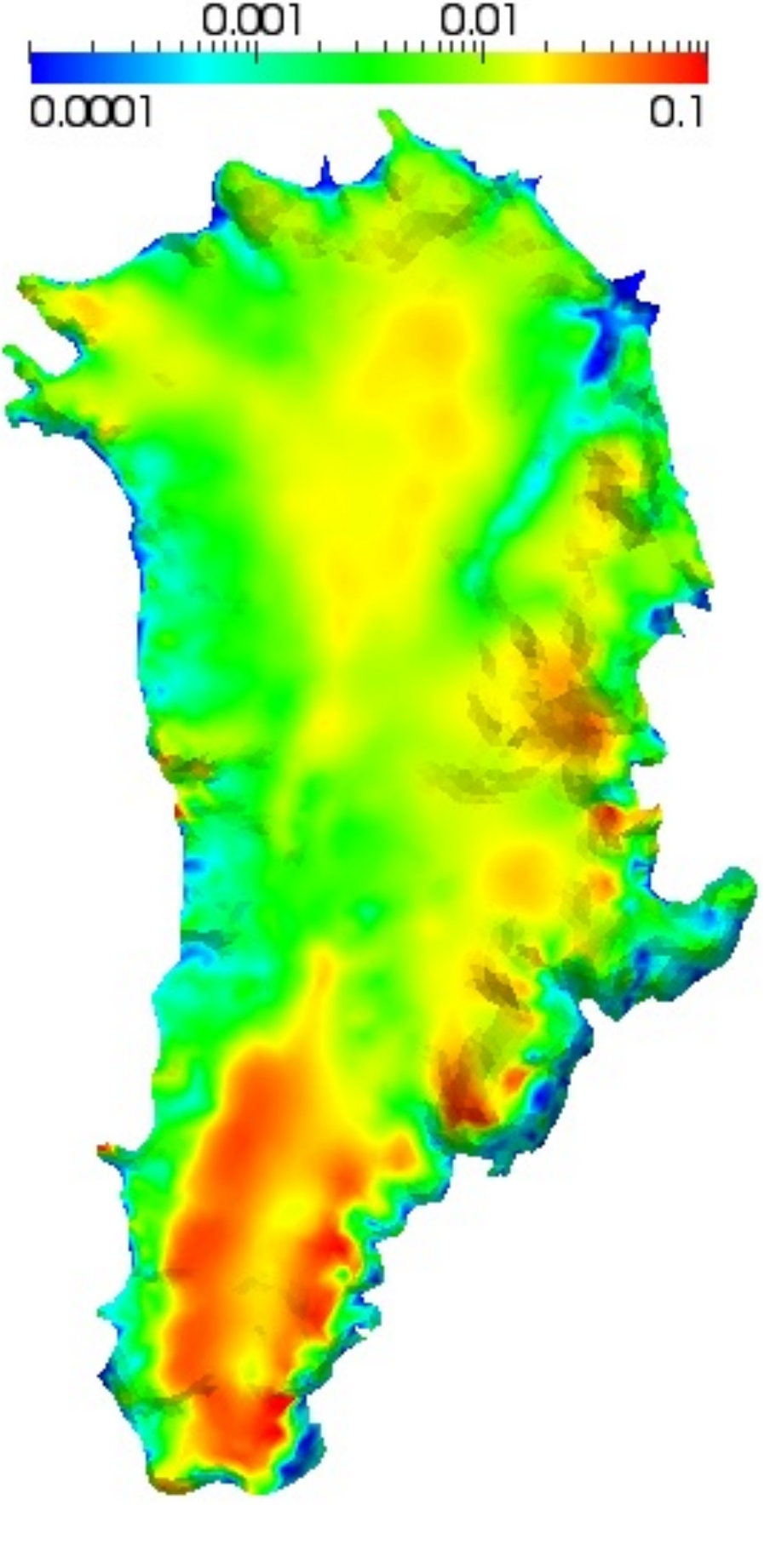} }\label{fig:geometrybeta}}%
    \caption{Experiment 4 - Input data for the simulation of the Greenland ice sheet. The geometry is from meausured data from \cite{Bamber}, and the sliding coefficient is computed in \cite{Fabien2012}. The data is smoothened to a resolution of $50$ km.}%
    \label{fig:geometry}
\end{figure}

Both the bedrock and the surface topography (\fgref{fig:geometry}) are from measurement data in \cite{Bamber}, while the sliding coefficient $\beta$ in \eqref{eq:slidinglaw} was computed by inverse modeling in \cite{Fabien2012}. Contrary to the model problem, the bedrock is far from flat, with the East Greenland mountain range, and large areas below sea level in the interior (\fgref{fig:geometryb}). The sliding coefficient is high (\textit{i.e.} high friction) in the south, and low in many outlet glaciers, \textit{e.g.} in the large ice stream in the northeast, NEGIS (Northeast Greenland Ice Stream) (\fgref{fig:geometrybeta}).

As the SIA is sensitive to high frequency spatial variations \cite{Hindmarsh,Ahlkrona13a}, the data are smoothed to a resolution of $50$ km (which is what is shown in \fgref{fig:geometry}). We generate a triangular mesh with an edge length of $10$--$20$ km. To simplify the mesh construction and avoid small skewed elements at bays and capes, we smooth the margin slightly.
The error estimation is based on the relative horizontal velocities, using a relative error tolerance as in Experiment 1 and 2, \textit{i.e.} $\epsilon_{rel}$ is $5$ \% and an absolute error tolerance, $\epsilon_{abs}$, of $1$ m/year.

\subsubsection{Results}
\begin{figure}
    \centering
    \subfloat[Velocity, ISCAL]{{\includegraphics[trim={6cm 10cm 1cm 2cm},clip,width=0.42\textwidth]{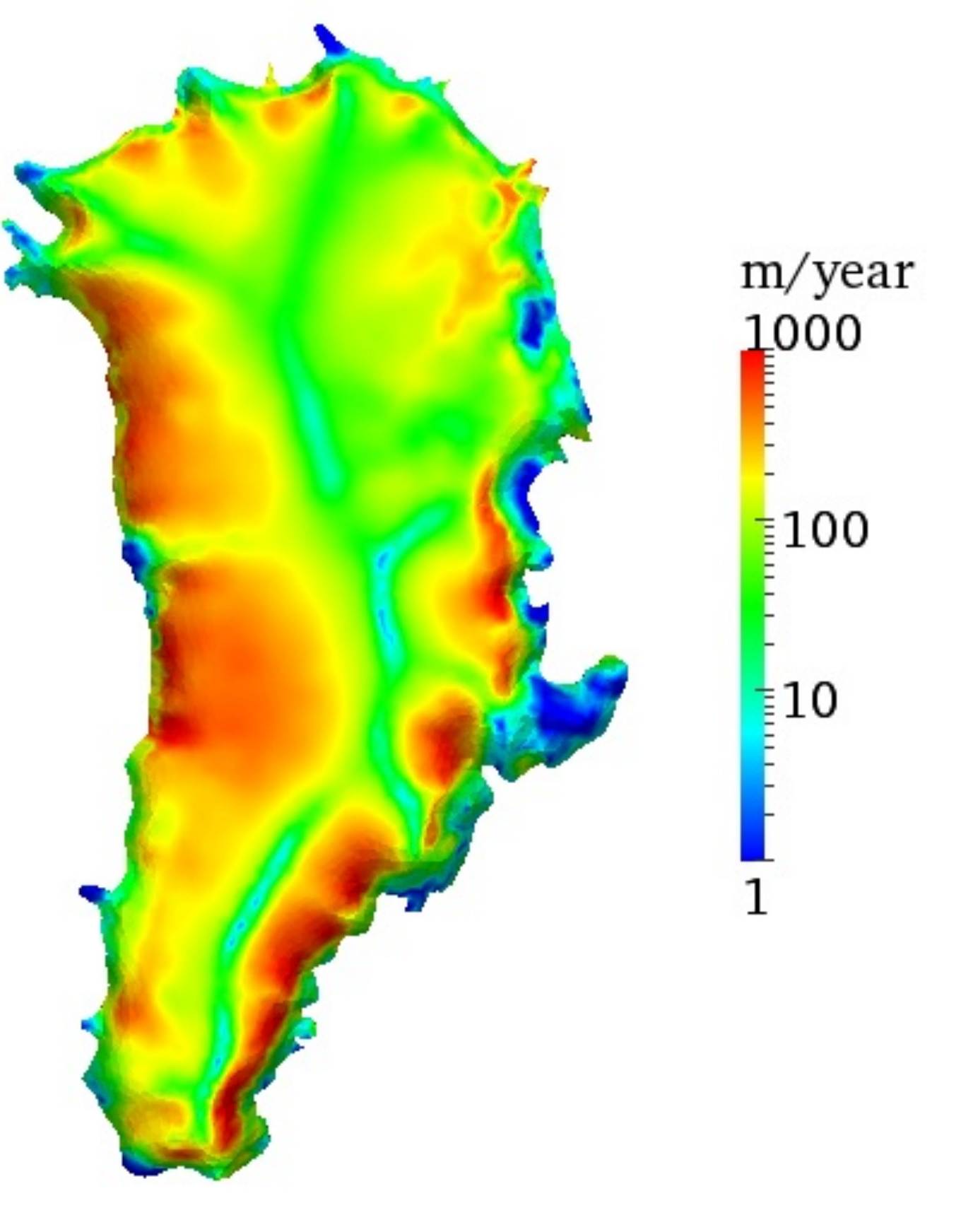} }\label{fig:greenlandveloa}}
    \qquad
    \subfloat[Observed velocity]{{\includegraphics[trim={13.5cm 3cm 6cm 2cm},clip,width=0.35\textwidth]{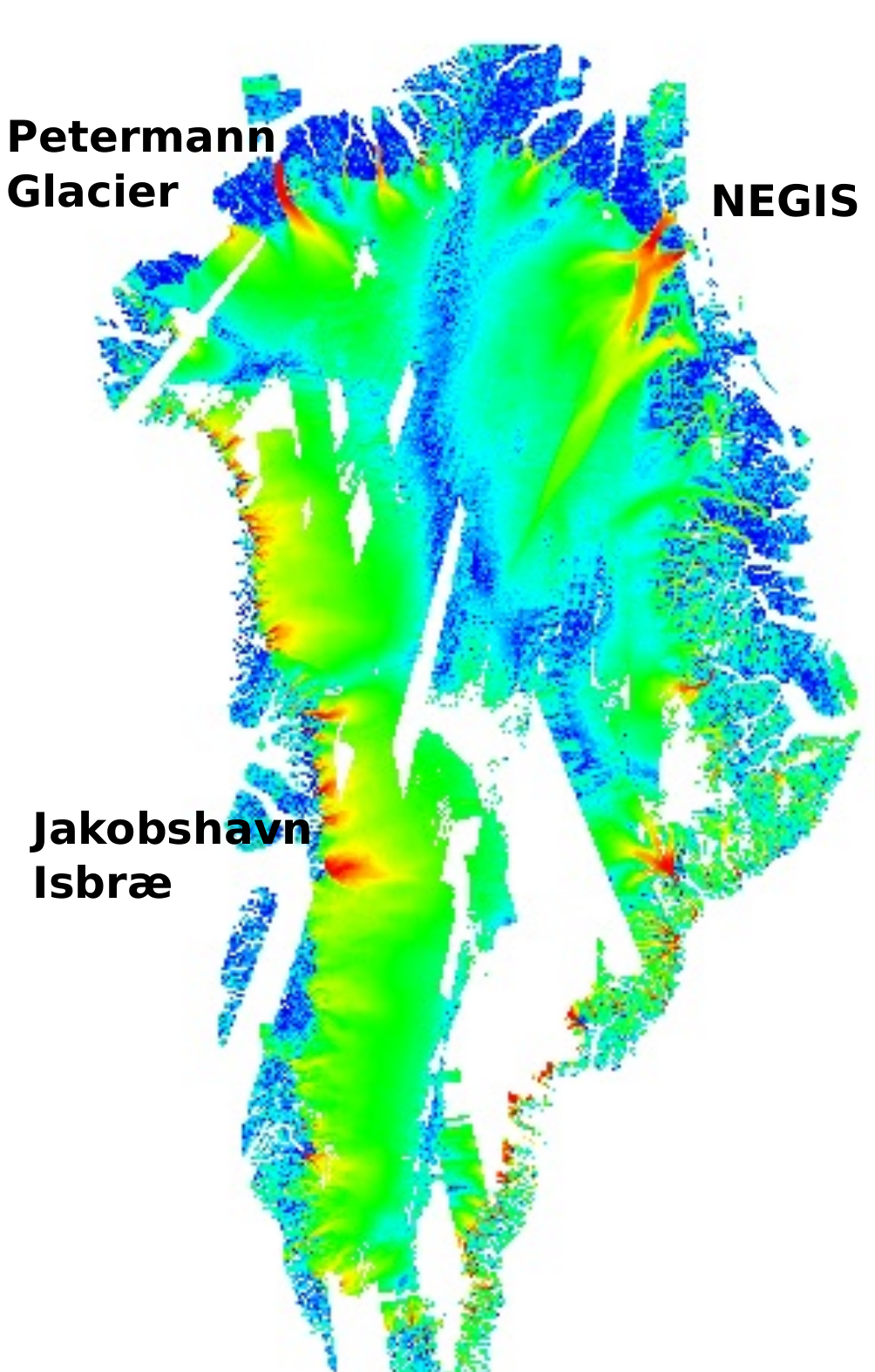} }\label{fig:greenlandvelob}}%
    \caption{Experiment 4 – ISCAL velocity and observed velocity (re-plotted from \cite{Ian}) and displayed on a common colorscale. White patches in the observed velocity are due to missing data. Three ice streams/outlet glaciers discussed in the text are named.}%
    \label{fig:greenlandvelo}
\end{figure}

The computed ISCAL velocity field is shown in \fgref{fig:greenlandveloa}, and compared with observational surface (horizontal) velocity data from \cite{Ian} (\fgref{fig:greenlandvelob}). As with the model problem, the velocity is high at the margins and in places with low friction, and low at the domes. The general pattern of the ISCAL velocity matches observations but the flow velocity is in general too high. The reason is because we did not allow the errors in the prescribed initial ice geometry to vanish, and since the data was smoothened and the mesh is fairly uniform. When simulating a real ice sheet, the velocity field and free surface are is usually allowed to relax, great care is taken into the initialization of the simulation. A realistic, transient, fully resolved, simulation of the Greenland Ice Sheet is however beyond the scope of this paper. The resolution is too low to capture all the small outlet glaciers at the margins, but large features such as the ice stream NEGIS in the northeast, the Petermann glacier in the northwest, and Jakobshavn Isbr{\ae} on the west coast are clearly visible.
\begin{figure}
    \centering
    \subfloat[Relative Error in SIA, $\Delta\uvec_{SIA}$]{{\includegraphics[trim={6cm 10cm 3cm 1.5cm},clip,width=0.34\textwidth]{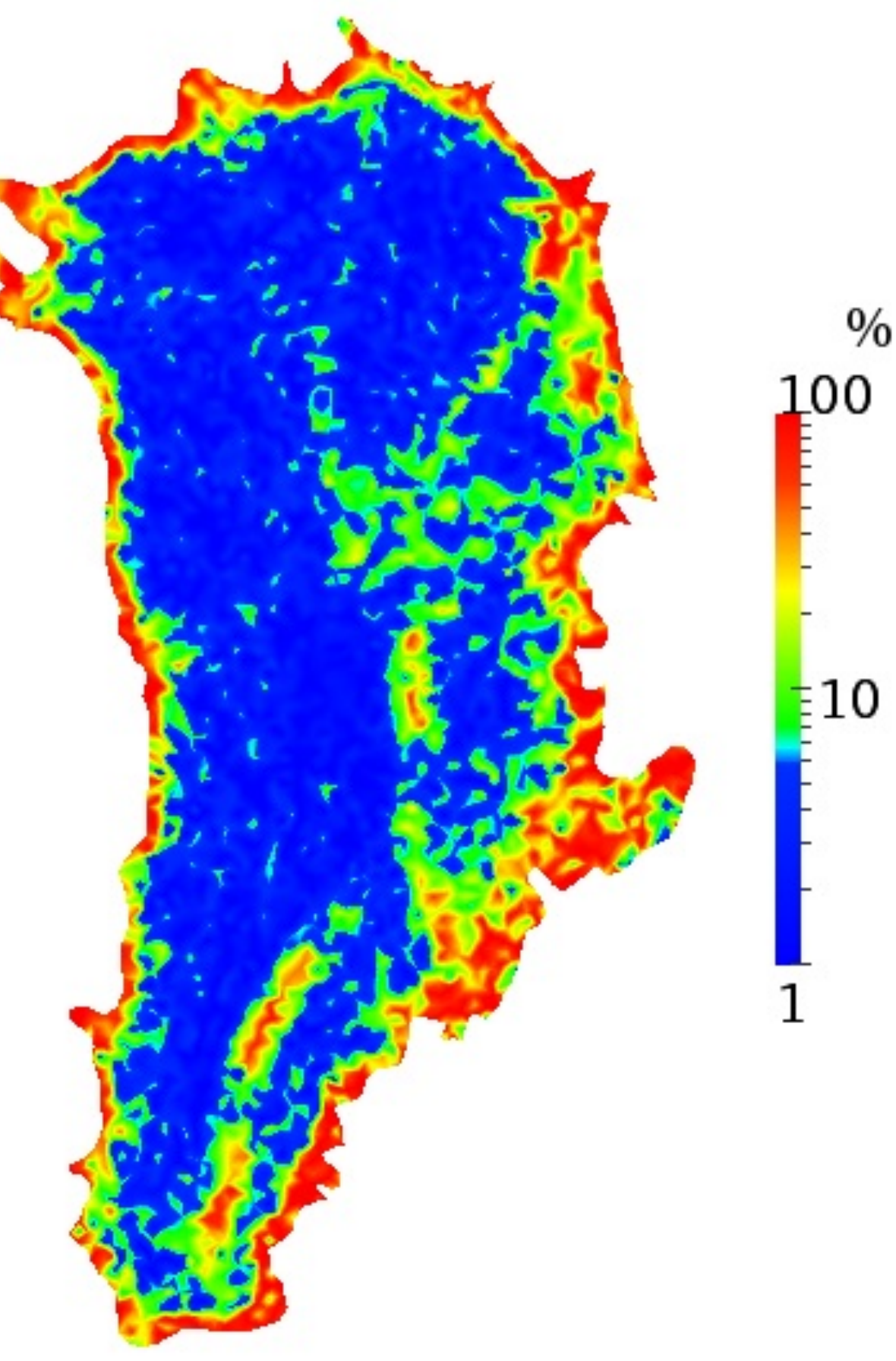} }\label{fig:greenlanderrora}}%
    \qquad
    \subfloat[$\Omega_{FS}$ (red) and $\Omega_{SIA}$ (blue)]{{\includegraphics[trim={6cm 10cm 6cm 1.5cm},clip,width=0.26\textwidth]{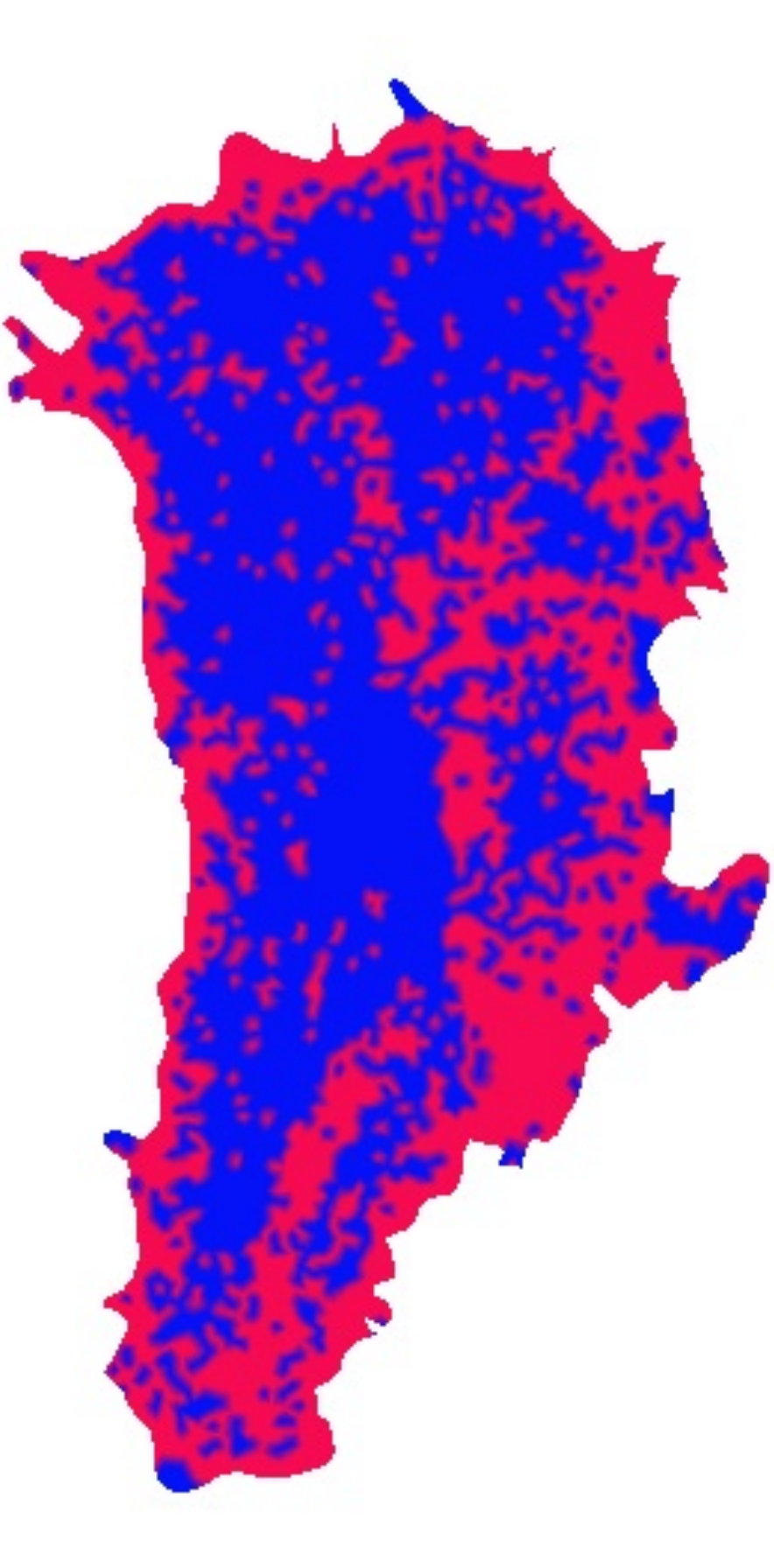} }\label{fig:greenlanderrorb}}%
    \caption{Experiment 4 - Estimated relative error of horizontal velocities in percent, and distribution of $\Omega_{FS}$ and $\Omega_{SIA}$. The relative error tolerance, $\epsilon_{rel}$, in SIA is $5$ \% and the absolute error tolerance, $\epsilon_{abs}$ is $1$ m/year.
}%
\label{fig:greenlanderror}
\end{figure}
The error in SIA in \fgref{fig:greenlanderrora} is high at the margins and domes (here ridges), \textit{cf.} the model problem in Section \ref{Efficiency}. Also in the high sliding area under the NEGIS in the northeast, the error is high. For this diagnostic simulation, the estimated SIA error and the true SIA error will be equal, as we always start a simulation with $\Omega_{FS}=\Omega$.  The distribution of $\Omega_{FS}$ and $\Omega_{SIA}$ in \fgref{fig:greenlanderrorb} follows from the error. As expected when using real-world data, the pattern is more irregular than for the model problem, but the general behavior is the same. The error tolerances are very low, and considering the uncertainty in data they can be allowed to be higher. If a relaxation of the velocity field and surface would have been applied, we expect the error and the extent of $\Omega_{FS}$ to decrease.

\section{Conclusions}\label{sec:concl}
ISCAL drastically reduces simulation time while keeping errors caused by the approximation of the Stokes equations at acceptable and controllable levels, by solving the full Stokes (FS) equations only where needed and using the SIA elsewhere. For a problem size of the same order of magnitude as for realistic ice sheet simulations, ISCAL is nine times as fast as FS, while the error in the horizontal velocity is no greater than $5$ \% or $1$ m/year. The speedup increases with problem size, mainly because the relative number of FS nodes is decreasing with increasing resolution. ISCAL automatically detects rapid changes in ice dynamics and applies the FS equations in regions where the SIA is not accurate enough, such as at margins, in ice streams, and at domes. The estimated error is close to the true error and also agrees with the theory of the SIA. Three different ways of determining the partial domain where to apply the SIA are implemented; estimating the error in the horizontal velocity, estimating the error through a functional of the velocity (ice flux), and estimating the error utilizing the residual of the Stokes equations. Each error estimate results in different distributions of the areas where the FS equations are solved, and are suitable for different applications. We further show that ISCAL is applicable to real problems, such as the Greenland Ice Sheet. The error and distribution of FS areas are more fragmented when using real, less smooth data, but the general pattern is the same and the SIA equations are sufficiently accurate in large parts of the ice sheet. 

When using the ISCAL method, special care should be given to the meshing procedure. As the SIA is sensitive to high frequency spatial variation in data (e.g. roughness in underlying bed), the data might have to be smoothened in order to avoid unnecessarily high SIA errors. Also, if the mesh is extra refined in areas where the SIA will likely not be valid, such as in ice streams, the speedup of the ISCAL will be lower. The ISCAL method is developed with paleosimulations in mind, i.e. large scale ice sheet simulations over long time spans. As the uncertainties in paleosimulations are anyway very large because of lack of accurate data, relatively coarse grids and smoothened data are not a problem.

Due to the computational costs of using the exact FS equations, the SIA and other approximations have long been the only option for paleosimulations, despite high errors in crucial areas. Considering how complex and variable the dynamics of a natural ice sheet is, it is obvious that any approximation will result in significant, over time accumulated, errors in some parts of the domain. We believe it is important not to apply approximations without controlling the modeling error, and ISCAL is a powerful tool for this.

\section{Acknowledgments}
Josefin Ahlkrona was supported by the Swedish strategic research programme eSSENCE, Nina Kirchner by the Bert Bolin Centre for Climate Research at Stockholm University, and Thomas Zwinger by the Nordic Centre of Excellence, eSTICC. We are grateful to Peter R{\aa}back and Juha Ruokolainen for advice and help in developing the ISCAL. We thank Fabien Gillet-Chaulet for providing the sliding parameter for Greenland, and Hakime Seddik for providing valuable instructions on how to use Elmer for real world applications. Further we thank the SeaRISE community (http://tinyurl.com/srise-umt) for the compilation of topography data sets. We also thank Evan Gowan and James Lea for help and for comments on the manuscript. The computations were performed on resources provided by the Swedish National Infrastructure for Computing (SNIC) at Uppmax at Uppsala University.


  \bibliographystyle{plain} 
  \bibliography{references}






\end{document}